\documentclass[a4paper,10pt]{article}
\usepackage[margin=1.2in]{geometry}

\usepackage{cancel,amsthm,amssymb,amsmath,mathrsfs,upgreek,stmaryrd}
\usepackage{ textcomp }
\usepackage{mathdots}
\usepackage{color,tikz}
\usetikzlibrary{patterns, decorations.markings,arrows, calc}
\usepackage{color,todonotes}
\usepackage{graphicx,framed,verbatim,caption}
\usepackage[colorlinks=true, pdfstartview=FitV, linkcolor=darkblue, citecolor=darkblue, urlcolor=darkblue]{hyperref}
\definecolor{shadecolor}{rgb}{0.9, 0.9, 0.86}
\definecolor{darkgreen}{rgb}{0.2, 0.5,  0}
\definecolor{darkblue}{rgb}{0.1,0.1,0.45}

\def\&{\vspace{-5pt}&}

\def\Re{\mathrm {Re}\,}

\def \wt{\widetilde}
\def \wh{\widehat}
\newcommand{\C}{\mathbb{C}}
\renewcommand{\le}{\left}
\newcommand{\ri}{\right}
\newcommand{\R}{\mathbb{R}}
\newcommand{\Z}{\mathbb{Z}}
\newcommand{\G}{\Gamma}
\newcommand{\1}{\mathbf{1}}

\renewcommand{\a}{\alpha}

\renewcommand{\L}{\Lambda}
\renewcommand{\l}{\lambda}
\renewcommand{\d}{\mathrm d}
\renewcommand{\O}{\mathcal{O}}
\renewcommand{\S}{\Sigma}
\newcommand{\U}{\mathrm{U}}
\renewcommand{\mod}{\,\mathrm{mod}\,}
\newcommand{\e}{\mathrm{e}}
\renewcommand{\i}{\mathrm{i}}
\renewcommand{\t}{\mathbf{t}}
\newcommand{\pa}{\partial}

\def\res{\mathop{\mathrm {res}}\limits_}
\def \tr {\mathrm{tr}\,}
\def\be{\begin{equation}}
\def\ee{\end{equation}}
\def\bg{\begin{gathered}}
\def\eg{\end{gathered}}

\def \U{\mathcal{U}}

\newtheorem{theorem}{Theorem}[section]

\newtheorem{lemma}[theorem]{Lemma}
\newtheorem{remark}[theorem]{Remark}
\newtheorem{problem}[theorem]{Riemann--Hilbert Problem}

\newtheorem{proposition}[theorem]{Proposition} 
\newtheorem{corollary}[theorem]{Corollary}

\def\QED {\hfill $\blacksquare$\par\vskip 3pt}
\def\s{\sigma}
\def\bea{\begin{eqnarray}}
\def\eea{\end{eqnarray}}

\DeclareMathOperator{\GL}{GL}

\DeclareMathOperator{\diag}{diag}

\makeatletter
\@addtoreset{equation}{section}
\makeatother

\begin{document}

\vspace{0.2cm}
\begin{center}
\begin{Large}
\textbf{Brezin--Gross--Witten tau function and isomonodromic deformations} 
\end{Large}
\end{center}

\bigskip
\begin{center}
M. Bertola$^{\dagger\ddagger \clubsuit}$\footnote{Marco.Bertola@\{concordia.ca, sissa.it\}},  
G. Ruzza $^{\ddagger}$ \footnote{giulio.ruzza@sissa.it}.
\\
\bigskip
\begin{minipage}{0.7\textwidth}
\begin{small}
\begin{enumerate}
\item [${\dagger}$] {\it  Department of Mathematics and
Statistics, Concordia University\\ 1455 de Maisonneuve W., Montr\'eal, Qu\'ebec,
Canada H3G 1M8} 
\item[${\ddagger}$] {\it SISSA, International School for Advanced Studies, via Bonomea 265, Trieste, Italy }
\item[${\clubsuit}$] {\it Centre de recherches math\'ematiques,
Universit\'e de Montr\'eal\\ C.~P.~6128, succ. centre ville, Montr\'eal,
Qu\'ebec, Canada H3C 3J7} 
\end{enumerate}
\end{small}
\end{minipage}
\vspace{0.5cm}
\end{center}
\bigskip

\begin{center}
\begin{abstract}
The Brezin--Gross--Witten tau function is a tau function of the KdV hierarchy which arises in the weak coupling phase of the Brezin--Gross--Witten model. It falls within the family of generalized Kontsevich matrix integrals, and its algebro--geometric interpretation has been unveiled  in recent works of Norbury. This tau function admits a natural extension, called generalized Brezin--Gross--Witten tau function. We prove that the latter is the isomonodromic tau function of a $2\times 2$ isomonodromic system and consequently present a study of this tau function purely by means of this isomonodromic interpretation. Within this approach we derive effective formul\ae\ for the generating functions of the correlators in terms of simple generating series, the Virasoro constraints, and discuss the relation with the Painlev\'{e} XXXIV hierarchy.
\end{abstract}


\end{center}

\setcounter{tocdepth}{2}

\tableofcontents

\section{Introduction and results}

The generalized Brezin--Gross--Witten (gBGW) tau function $\tau(\t;\nu)$ is a formal tau function of the Korteweg--de Vries (KdV) hierarchy; it depends on infinitely many ``times'' $\t=(t_0,t_1,t_2,...)$ which  are the usual flows of the KdV hierarchy, while the  parameter $\nu\in\Z$ plays the role of an additional discrete time of the hierarchy. With respect to the $\nu$--dependence it is a tau function of the \emph{modified Kadomtsev--Petviashvili} hierarchy \cite{Al2016}. The restriction $\nu=0$ corresponds to the what is usually called BGW tau function.

This tau function arises in the weak coupling phase of the BGW model \cite{GrWi1980,BrGr1980} and was studied in \cite{GrNe1992,MiMoSe1996,Al2016,DoNo2016}; we review the definition of $\tau(\t;\nu)$ along with its relation with the BGW model in Sec. \ref{secdeftau} below.

The first few terms of its formal expansion read
\begin{align}\label{tau}
&\tau(\t;\nu)=1+\frac{1-4\nu^2}{16}t_0+
\frac{(1-4\nu^2)(9-4\nu^2)}{1024}(t_1+2t_0^2)
\\
&\qquad+\frac{(1-4\nu^2)(9-4\nu^2)(25-4\nu^2)}{32768}(t_2+2t_0t_1)+
\frac{(1-4\nu^2)(9-4\nu^2)(17-4\nu^2)}{24576}t_0^3+\cdots.\nonumber
\end{align}

In \cite{No2017} the author has found the algebro--geometric interpretation of $\tau(\t;\nu=0)$ (i.e. the BGW tau function proper) as a generating function of intersection numbers on the moduli spaces $\overline{\mathcal{M}_{g,n}}$ of stable curves of genus $g$ with $n$ marked points, a result which parallels the Witten--Kontsevich Theorem \cite{Wi1991,Ko1992}. More precisely, in \cite{No2017} the author constructed certain cohomology classes $\Theta_{g,n}\in H^{2(2g-2+n)}\left(\overline{\mathcal{M}_{g,n}};\mathbb{Q}\right)$ for all $g,n\geq 0$ such that $2g-2+n\geq 1$. He also proved that
\begin{align}\label{1}
\log\tau(\t;\nu=0)&=\frac{1}{16}t_0+\frac{9}{1024}t_1+\frac{1}{64}t_0^2+
\frac{225}{32768}t_2+\frac{27}{2048}t_0t_1+\frac{1}{192}t_0^3+\cdots
\\ &=\sum_{g\geq 0}\sum_{n\geq 0}\sum_{\ell_1,...,\ell_n\geq 0}\frac{1}{n!}\left(\prod_{j=1}^n\frac{(2\ell_j+1)!!}{2^{2\ell_j+1}}t_{\ell_j}\right)\int_{\overline{\mathcal{M}_{g,n}}}\Theta_{g,n} \psi_1^{\ell_1} \cdots \psi_n^{\ell_n},\nonumber
\end{align}
where $\psi_j\in H^2\left(\overline{\mathcal{M}_{g,n}};\mathbb{Q}\right)$ is, as customary, the first Chern class of the cotangent
line bundle at the $j$--th marked point, $j=1,...,n$; the dimensional constraint implies $g=\ell_1+\cdots+\ell_n+1$ in \eqref{1}.

Conjecturally \cite{Al2016,AlBuTe2017} the $\nu$--dependence of $\tau(\t;\nu)$  should encode some deformation of the intersection numbers constructed in \cite{No2017}.

The main aim of this paper is to interpret the gBGW tau function as an isomonodromic tau function, see details below. This isomonodromic approach allows us to explicitly compute all these intersection numbers by means of the formul\ae\ of Thm. \ref{thmcor} below.

To state the theorem, let us introduce the generating functions $S_n(z_1,...,z_n;\nu)$, for $n\geq 1$,
\be\label{defsn}
S_n(z_1,...,z_n;\nu):=\sum_{\ell_1,...,\ell_n\geq 0}\frac{1}{z^{1+\ell_1}_1\cdots z_n^{1+\ell_n}}\left.\frac{\pa^n\tau(\t;\nu)}{\pa t_{\ell_1}\cdots\pa t_{\ell_n}}\right|_{\t=0}
\ee
and the matrix $\mathcal{U}(z;\nu)$
\be\label{U}
\renewcommand*{\arraystretch}{2}
\mathcal{U}(z;\nu):=\sum_{k\geq 0}\frac{(2k-1)!!}{k!(8z)^k}
\begin{bmatrix}
\frac{1}{2}\left(\frac 1 2-\nu\right)_{k+1}\left(\frac 1 2+\nu\right)_k
&
\left(\frac 1 2-\nu\right)_k\left(\frac 1 2+\nu\right)_k
\\
-z\left(\frac 1 2-\nu\right)_{k+1}\left(\frac 1 2+\nu\right)_{k-1}
 &
-\frac{1}{2}\left(\frac 1 2-\nu\right)_{k+1}\left(\frac 1 2+\nu\right)_k
\end{bmatrix}
\ee
where hereafter $\left(\a\right)_\ell:=\a(\a+1)\cdots(\a+\ell-1)$ denotes the rising factorial, and conventionally we set $(\a)_0:=1$ and $(\a)_{-1}:=\frac{1}{\a-1}$; we also agree that $(-1)!!:=1$.
Then the main theorem can be stated as follows:
\begin{shaded}
\begin{theorem}\label{thmcor}
For all $\ell\geq 0$ we have
\be\label{onepoint}
\left.\frac{\pa\tau(\t;\nu)}{\pa t_\ell}\right|_{\t=0}= \frac{(2\ell-1)!!}{2^{3\ell+2}(\ell+1)!}\left(\frac 1 2 -\nu\right)_{\ell+1} \left(\frac 1 2 +\nu\right)_{\ell+1}
\ee
and for all $n\geq 2$ we have
\be\label{npoint}
S_n(z_1,...,z_n;\nu)=\frac{(-1)^{n-1}}{n}\sum_{\iota\in \mathfrak{S}_n}\frac{\tr\left(\mathcal{U}(z_{\iota_1};\nu)\cdots\mathcal{U}(z_{\iota_n};\nu)\right)}{(z_{\iota_1}-z_{\iota_2})\cdots(z_{\iota_{n-1}}-z_{\iota_n})(z_{\iota_n}-z_{\iota_1})}-\frac{z_1+z_2}{(z_1-z_2)^2}\delta_{n,2}.
\ee
\end{theorem}
\end{shaded}

\noindent Thm. \ref{thmcor} is proven in Sec. \ref{secproofcor}. Note that $\mathcal{U}(z;\nu)$ is a power series in $z$ whose coefficients are polynomials in $\nu$. Moreover, $\U(z;\nu)$ satisfies the following identity
\be
\U(z;-\nu)=\begin{bmatrix}
1 & 0 \\ -\nu & 1 
\end{bmatrix}
\U(z;\nu)
\begin{bmatrix}
1 & 0 \\ \nu & 1 
\end{bmatrix}
\ee
from which we conclude, using \eqref{npoint}, that the gBGW tau function is invariant under $\nu\mapsto-\nu$, namely all the coefficients in the expansion of the gBGW tau function are even polynomials in $\nu$.

In particular when $\nu$ is a half--integer, $\U(z;\nu)$ is actually a Laurent polynomial in $z$ which reflects the fact that the gBGW tau function is a polynomial in this case; see \cite{Al2016} for a description of these polynomials in terms of Schur polynomials.

As an application of Thm. \ref{thmcor} we can derive explicit formul\ae\ for the intersection numbers of \cite{No2017} by setting $\nu=0$; more precisely, identifying
\be 
\int_{\overline{\mathcal{M}_{g,n}}}\Theta_{g,n} \psi_1^{\ell_1} \cdots \psi_n^{\ell_n}=\frac{2^{2\ell_1+1}\cdots 2^{2\ell_n+1}}{(2\ell_1+1)!!\cdots(2\ell_n+1)!!}\left.\frac{\pa^n\tau(\t;\nu=0)}{\pa t_{\ell_1}\cdots \pa t_{\ell_n}}\right|_{\t=0}
\ee
from \eqref{1}, we have the following immediate Corollary.

\begin{shaded}
\begin{corollary}\label{corcor}
For all $g\geq 1$ we have
\be\label{onepointnorb}
\int_{\overline{\mathcal{M}_{g,1}}}\Theta_{g,1} \psi_1^{g-1}=\frac{(2g-1)!!(2g-3)!!}{8^gg!}
\ee
and for all $n\geq 2$ we have
\begin{align}
&\sum_{\ell_1,...,\ell_n\geq 0}\frac{(2\ell_1+1)!!\cdots (2\ell_n+1)!!}{2^{2\ell_1+1}\cdots 2^{2\ell_n+1}z^{1+\ell_1}_1\cdots z_n^{1+\ell_n}}\int_{\overline{\mathcal{M}_{g,n}}}\Theta_{g,n} \psi_1^{\ell_1} \cdots \psi_n^{\ell_n}
\\
&\qquad =\frac{(-1)^{n-1}}{n}\sum_{\iota\in \mathfrak{S}_n}\frac{\tr\left(\mathcal{U}(z_{\iota_1};\nu=0)\cdots\mathcal{U}(z_{\iota_n};\nu=0)\right)}{(z_{\iota_1}-z_{\iota_2})\cdots(z_{\iota_{n-1}}-z_{\iota_n})(z_{\iota_n}-z_{\iota_1})}-\frac{z_1+z_2}{(z_1-z_2)^2}\delta_{n,2}.\nonumber
\end{align}
\end{corollary}
\end{shaded}

With the aid of these formul\ae\ we have computed several intersection numbers reported in the tables of App. \ref{table}.

\begin{remark}
From \eqref{onepointnorb} we can write a closed form for the generating function of the one--point intersection numbers as follows 
\begin{shaded}
\be\label{tricomi}
\sum_{g\geq 1}X^{g}\int_{\overline{\mathcal{M}_{g,1}}}\Theta_{g,1} \psi_1^{g-1}=\frac{X}{8}+\frac{3X^2}{128}+\frac{15X^3}{1024}+\frac{525 X^4}{32768}+\cdots\sim 1+\i\sqrt{\frac{X}{2}}\, \mathrm{U}\left(-\frac{1}{2},0,-\frac{2}{X}\right)
\ee
\end{shaded}
\noindent where $\mathrm{U}(a,b,z)$ is the Tricomi confluent hypergeometric function \cite{AbSt1965}, and symbol $\sim$ denotes the equality as asymptotic expansion, which here is valid as $X\to 0$ within the sector $\Re X>0$.
\end{remark}

The identification of the Brezin-Gross-Witten tau function as an appropriate isomonodromic tau function allows us also to derive independently the Virasoro constraints for this model, already known in the case $\nu=0$ from \cite{GrNe1992,MiMoSe1996,DoNo2016} and in the general case from \cite{Al2016} by other methods. In concrete terms, we introduce the following differential operators;
\be\label{vir}
L_m:=\sum_{\ell\geq 0}\frac{2\ell+1}{2}\left(t_\ell-2\delta_{\ell,0}\right)\frac{\pa}{\pa t_{\ell+m}}+\frac{1}{4}\sum_{\ell=0}^{m-1}\frac{\pa^2}{\pa t_{\ell}\pa t_{m-1-\ell}}+\left(\frac{1-4\nu^2}{16}\right)\delta_{m,0}, \ \ \ m\geq 0.
\ee
They satisfy the Virasoro commutation relations
\be
\label{virasorocomm}
[L_m,L_n]=(m-n)L_{m+n}, \ \ \ m,n\geq 0.
\ee
 
\begin{shaded}
\begin{theorem}[\cite{Al2016}]
\label{thmvir}
The Virasoro operators annihilate the gBGW tau function;
\be
L_m\tau(\t;\nu)=0, \qquad m\geq 0.
\ee
\end{theorem}
\end{shaded}

\noindent The proof of Thm. \ref{thmvir} by the isomonodromic method is contained in Sec. \ref{secproofvir}. Note that the situation is slightly different from the Witten--Kontsevich case, where the Virasoro constraints include an additional equation $L_{-1}\tau =0$ \cite{Wi1991}.

Below we provide details on the approach and on the main results; proofs are deferred to Sec. \ref{secproof}.

\subsection{The Brezin--Gross--Witten tau function}\label{secdeftau}

We consider a partition function \cite{GrWi1980,BrGr1980} given by the following unitary matrix integral
\be\label{bgwpartition}
\wh Z_n(\L;\nu):=\int_{\mathrm{U}_n}\frac{{\det}^\nu J}{{\det}^\nu U}\exp\tr \frac 1 \beta \left( J^\dagger U+JU^\dagger\right) \d U, \qquad \L:=\frac 1 \beta \left(JJ^\dagger\right)^\frac 1 2
\ee
where $\d U$ denotes the normalized Haar measure on the unitary group $\mathrm{U}_n$, $\int_{\mathrm{U}_n}\d U=1$. The parameter $\beta$ is the coupling constant and the external field $J$ is a complex $n\times n$ matrix; however, as emphasized in the notation $\wh Z_n(\L;\nu)$, the partition function \eqref{bgwpartition} actually depends only on the eigenvalues of the Hermitian matrix $\L$ defined in \eqref{bgwpartition}. Without loss of generality we are going to assume that $\L$ is diagonal with eigenvalues $\l_1,...,\l_n$ and that $\beta=1$.

The parameter $\nu$ in \eqref{bgwpartition} was absent in the original formulation of the model and is added here to match with the generalization introduced in \cite{MiMoSe1996,Al2016}. Interestingly, this type of generalization had appeared also in the Physics literature on QCD, see e.g. \cite{LeSm1992,JaSeVe1996,AkWa1998}.

It was first argued in \cite{MiMoSe1996} that $\wh Z_n(\L;\nu)$ can be identified with a generalized Kontsevich model \cite{KhMaMiMoZa1992} with non--polynomial potential $M^{-1}+\nu\log M$, see \eqref{generalizedKontdef} below. 
We now describe this relationship in detail. 

First, by a character expansion it is possible to compute \cite{Ba2000,ShWe2003}
\be\bg\label{evviva}
\wh Z_n(\L;\nu)=
\prod_{j=1}^{n-1}j!\ \frac{\det [\l_j^{k+\nu-1}I_{k-\nu-1}(2\l_j)]_{j,k=1}^n}{\Delta(\l_1^2,...,\l_n^2)}
\eg\ee
where $I_{\a}(x)$ denotes the modified Bessel functions of the first kind of order $\a$ \cite{AbSt1965}, and $\Delta(\xi_1,...,\xi_n):=\det[\xi_j^{k-1}]_{j,k=1}^n=\prod\limits_{i<j}(\xi_j-\xi_i)$ denotes the Vandermonde determinant. 

Introduce now the following generalized Kontsevich matrix integral \cite{KhMaMiMoZa1992,MiMoSe1996,Al2016}
\be\label{generalizedKontdef}
Z_n(\L;\nu):=\frac{\int_{\mathrm{H}_n(\gamma)}\exp\tr \left(\L^2 M+M^{-1}+(\nu-n)\log M\right)\d M}{\int_{\mathrm{H}_n(\gamma)}\exp\tr \left(M^{-1}+(\nu-n)\log M\right)\d M}
\ee
where $\mathrm{H}_n(\gamma):=\{M=U \diag(x_1,...,x_n) U^\dagger: \ U\in\mathrm{U}_n, \ x_j\in\gamma\}$, $\gamma$ being a contour from $-\infty$ encircling zero counterclockwise once and going back to $-\infty$. With the help of the Harish--Chandra--Itzykson--Zuber formula one can show that
\be  \label{ecometichiamo}
Z_n(\L;\nu)=\prod_{j=1}^n\G(j-\nu)\ \frac{\det[\l^{k+\nu-1}_jI_{k-\nu-1}(2\l_j)]_{j,k=1}^n}{\Delta(\l_1^2,...,\l_n^2)}.
\ee
Comparing \eqref{ecometichiamo} with \eqref{evviva} we finally conclude that
\be\label{generalizedKont}
\wh Z_n(\L;\nu)=\prod_{j=1}^n\frac{\G(j)}{\G(j-\nu)}\ Z_n(\L;\nu).
\ee

\begin{remark}
In \eqref{bgwpartition} $\nu$ must be an integer, as the function ${\det}^\nu U$ is otherwise multi-valued on $\mathrm{U}_n$ and the integral makes no sense. Nonetheless in \eqref{generalizedKontdef} $\nu$ can be any complex number such that $\nu\not=1,2,3...$; notice however that such poles come from the normalizing denominator in \eqref{generalizedKontdef} only. 
\end{remark}

In the large $\L$ limit, corresponding to the weak coupling phase $\beta\to 0$ in \eqref{bgwpartition}, we consider the following expression \cite{Al2016}
\be
\label{taun}
\tau_n(\l_1,...,\l_n;\nu):=\frac{(2\pi)^{\frac n 2}\prod_{i,j=1}^n\sqrt{\l_i+\l_j}}{\e^{2\tr \L}{\det}^\nu\L\prod_{j=1}^{n}\G(j-\nu)}Z_n(\L;\nu)=\frac{\det[2\sqrt{\pi\l_j}\e^{-2\l_j}\l_j^{k-1}I_{k-\nu-1}(2\l_j)]_{j,k=1}^n}{\Delta(\l_1,...,\l_n)}
\ee
which admits a regular asymptotic expansion as $|\l_j|\to \infty$ within the sector $|\arg\l_j|<\frac{\pi}{2}-\delta$ for all $j=1,...,n$; this is easily seen because the Bessel functions have the following regular asymptotic expansion \footnote{I.e. an asymptotic expansion in integer powers of $\l$ only, e.g. without exponential factors.}
\be\label{phij}
2\sqrt{\pi\l} \e^{-2\l} I_{\a}(2\l)\sim 1+\O\left(\l^{-1}\right)
\ee
as $\l\to\infty$ within the sector $|\arg\l|\leq\frac{\pi}{2}-\delta$, for any $\delta>0$ \cite{AbSt1965}.
It is known that such an expansion for large $\L$ can be written as $n\to\infty$ as a formal power series in the odd Miwa variables
\be 
\label{miwa}
t_\ell(\l_1,...,\l_n):=\frac{\l^{-2\ell-1}_1+\cdots+\l^{-2\ell-1}_n}{2\ell+1}, \ \ \ell\geq 0.
\ee
The gBGW tau function is, by definition, the formal expansion of \eqref{taun} for large $\L$ written in terms of the Miwa times \eqref{miwa}.
The limit $n\to\infty$ means that the expansion of \eqref{taun} is a symmetric formal series in $\l_1^{-1},...,\l_n^{-1}$, which can therefore be expressed in terms of the symmetric polynomials $p_k=k^{-1}\sum\l_j^{-k}$; the coefficients in front of any monomial in the $p$'s then stabilize for $n\to\infty$ and vanish for monomials involving even $p$'s. A complete proof of these statements can be extracted from \cite{ItZu1992} or \cite[Chap. 14]{Di2003}.

The determinantal representation \eqref{taun} and its subsequent generalization \eqref{taunext} below are the starting point of our further considerations.

\subsection{The bare ODE}

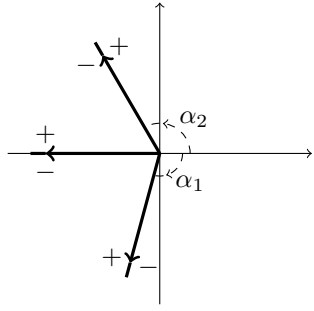
\begin{figure}[htbp]
\centering
\hspace{1cm}
\begin{tikzpicture}

\draw[->] (-2,0) -- (2,0);
\draw[->] (0,-2) -- (0,2);

\begin{scope}[rotate=-105]
\node at (1.5,.25) {$-$};
\node at (1.5,-.25) {$+$};
\draw[->,very thick] (0,0) -- (1.5,0);
\draw[very thick] (1.5,0) -- (1.7,0);
\end{scope}

\begin{scope}[rotate=120]
\node at (1.5,.25) {$-$};
\node at (1.5,-.25) {$+$};
\draw[->,very thick] (0,0) -- (1.5,0);
\draw[very thick] (1.5,0) -- (1.7,0);
\end{scope}

\begin{scope}[rotate=180]
\node at (1.5,.25) {$-$};
\node at (1.5,-.25) {$+$};
\draw[->,very thick] (0,0) -- (1.5,0);
\draw[very thick] (1.5,0) -- (1.7,0);
\end{scope}
\draw[dashed, decoration={markings, mark=at position 0.7 with {\arrow{>}}}, postaction={decorate}] (.3,0) arc (0:-105:.3);
\node at (.4,-.4) {$\a_1$};
\draw[dashed, decoration={markings, mark=at position 0.7 with {\arrow{>}}}, postaction={decorate}] (.4,0) arc (0:120:.4);
\node at (.45,.45) {$\a_2$};
\end{tikzpicture}
\caption{Contour $\Sigma$, and notation for the boundary values.}
\label{fig1}
\end{figure}

The strategy of our proof involves the dressing of a bare Riemann Hilbert problem; this is the Riemann--Hilbert problem induced by the Stokes' phenomenon of a linear ODE in the complex plane, which we refer to as the ``bare ODE''. To formulate this bare problem  we fix two angles $\a_1$, $\a_2$ in the range
\be
\label{alpha} -\pi<\a_1<\a_2<\pi
\ee 
and define $\S$ to be the contour in the $z$--plane consisting of the three rays $z<0$, $\arg z=\a_1$, $\arg z=\a_2$, see Fig. \ref{fig1}. Introduce the following $2\times 2$ matrix $\Xi(z)$, analytic for $z\in\C\setminus\S$:

\be\label{Psi0}
\Xi(z):=\sqrt{\frac{2}{\pi}}\times \begin{cases}
 \begin{bmatrix}
\pi I_{-\nu}(2\sqrt{z})+\i\e^{\i\nu\pi} K_{-\nu}(2\sqrt{z}) & -K_{-\nu}(2\sqrt{z}) \\
\pi \sqrt{ z}I_{1-\nu}(2\sqrt{z})-\i\e^{\i\nu\pi}\sqrt{z}K_{1-\nu}(2\sqrt{z}) & \sqrt{z} K_{1-\nu}(2\sqrt{z})
\end{bmatrix}& -\pi<\arg z<\a_1
\\
\\
\begin{bmatrix}
\pi I_{-\nu}(2\sqrt{z}) & -K_{-\nu}(2\sqrt{z}) \\
\pi\sqrt{ z}I_{1-\nu}(2\sqrt{z}) & \sqrt{z}K_{1-\nu}(2\sqrt{z})
\end{bmatrix}& \a_1<\arg z<\a_2
\\
\\
\begin{bmatrix}
\pi I_{-\nu}(2\sqrt{z})-\i \e^{-\i\nu\pi}K_{-\nu}(2\sqrt{z}) & -K_{-\nu}(2\sqrt{z}) \\
\pi\sqrt{ z}I_{1-\nu}(2\sqrt{z})+\i\e^{-\i\nu\pi}\sqrt{z}K_{1-\nu}(2\sqrt{z}) & \sqrt{z}K_{1-\nu}(2\sqrt{z})
\end{bmatrix}& \a_2<\arg z<\pi
\end{cases}
\ee
where $I_\a(x)$, $K_\a(x)$ are the modified Bessel functions of order $\a$ of the first and second kind respectively \cite{AbSt1965} and we stipulate henceforth that all the roots are principal. Note that we are implying the dependence on $\nu$.

The following proposition is elementary and the proof is omitted.

\begin{proposition}
In every sector of $\C\setminus\S$ the following statements hold true.

\begin{enumerate}
\item The following ODE is satisfied;\footnote{Hereafter we denote $'=\frac{\d}{\d z}$.}
\be\label{bare}
\renewcommand*{\arraystretch}{1.25}
\Xi'(z)=\begin{bmatrix}
-\frac{\nu}{2z}&\frac{1}{z} \\ 1 &\frac{\nu}{2z}
\end{bmatrix}\Xi(z).
\ee
\item We have the asymptotic expansion below;\footnote{We use the Pauli matrices $\s_1=\begin{bmatrix}
0&1 \\ 1 &0 \end{bmatrix}$ and $\s_3=\begin{bmatrix}
1&0 \\ 0 &-1 \end{bmatrix}$.}
\be\label{aaasymp}
\Xi(z)\sim z^{-\frac{\s_3}{4}}G\left(\1+\frac{1}{16\sqrt z}\begin{bmatrix}
-(1-2\nu)^2 & 2-4\nu\\ -2+4\nu & (1-2\nu)^2
\end{bmatrix}
+\O\left(z^{-1}\right)\right)\e^{2\sqrt{z}\s_3}, \ \ z\to \infty
\ee
where
\be\label{G}
G:=\frac{ 1}{ \sqrt{2}} \begin{bmatrix}
1 & - 1 \\ 1 & 1 \end{bmatrix}.
\ee
\item We have $\det\Xi(z)\equiv 1$.
\end{enumerate}
\end{proposition}

\noindent Moreover, the matrix $\Xi(z)$ satisfies the following jump condition along $\S$;
\be
\Xi(z_+)=\Xi(z_-) S(z), \ \ z\in\S
\ee
where $\pm$ denote boundary values as in Fig. \ref{fig1} and $S(z)$ is the following piecewise constant matrix defined on $\S$;
\be\label{stokes}
S(z):=\begin{cases}\i\s_1 & z<0 \\
\\
\begin{bmatrix}
1&0 \\ -\i\e^{\i\nu\pi} & 1
\end{bmatrix}& \arg z=\a_1 \\
\\
\begin{bmatrix}
1&0 \\ -\i\e^{-\i\nu\pi} & 1
\end{bmatrix}& \arg z=\a_2. \\
\end{cases}
\ee

\subsection{Extension of $\tau_n(\l_1,...,\l_n;\nu)$}

For later convenience we introduce an extension of $\tau_n(\l_1,...,\l_n;\nu)$, defined in \eqref{taun}, having the same regular asymptotic expansion when the $\l_j$'s go to infinity within arbitrary sectors of the $\l$-plane, not only within a sector $|\arg\l_j|<\frac{\pi}{2}-\delta$ for any $\delta>0$, as for \eqref{taun}. The strategy is parallel to that of \cite{BeCa2017}.

We introduce, for $-\pi<\arg\l<\pi$ and $k\geq 1$, the functions
\be
\label{defxik}
\xi_k(\l):=\sqrt{\frac{2}{\pi}}\l^{k-1}\times
\begin{cases}
\i K_{k-\nu-1}(2\e^{\i\pi}\l) & \text{if }-\pi<\arg\l<-\frac\pi 2 \\
\pi I_{k-\nu-1}(2\l)-\i\e^{\i(k-\nu)\pi} K_{k-\nu-1}(2\l) & \text{if }-\frac{\pi}{2}<\arg\l<\frac{\a_1}{2}\\
\pi I_{k-\nu-1}(2\l) & \text{if }\frac{\a_1}{2}<\arg\l<\frac{\a_2}{2}\\
\pi I_{k-\nu-1}(2\l)+\i\e^{\i(k+\nu)\pi} K_{k-\nu-1}(2\l) & \text{if }\frac{\a_2}{2}<\arg\l<\frac{\pi}{2}\\
-\i K_{k-\nu-1}(2\e^{-\i\pi}\l) & \text{if }\frac \pi 2 <\arg\l<\pi.
\end{cases}
\ee
The motivation behind this convoluted definition is that the above functions have the {\it same} asymptotic expansion
\be\label{asymptoticxi}
\xi_k(\l)\sim\frac{ 1}{ \sqrt{2\l}} \e^{2\l}\l^{k-1}\left(1+\O\left(\l^{-1}\right)\right), \ \ \l\to\infty
\ee
in every sector of $-\pi<\arg\l<\pi$ appearing in the definition \eqref{defxik}.

\begin{remark}\label{remrem}
Note that
\be\bg
\label{eqxi}
\xi_1(\l)=\begin{cases}
\Xi_{11}(\l^2) & \text{if }-\frac\pi 2<\arg\l<\frac\pi 2
\\
\pm\i\Xi_{12}(\l^2\e^{\mp 2\pi\i}) & \text{if }\frac \pi 2 <\pm \arg\l<\pi
\end{cases}
\\
\xi_2(\l)=
\begin{cases}
\Xi_{21}(\l^2) & \text{if }-\frac\pi 2<\arg\l<\frac\pi 2
\\
\mp\i\Xi_{22}(\l^2\e^{\mp 2\pi\i}) & \text{if }\frac \pi 2 <\pm\arg\l<\pi.
\end{cases}
\eg\ee
\end{remark} 

For arbitrary $\l_1,...,\l_n$ in $\C\setminus \S$ \footnote{We are free to deform the contour $\S$ if necessary, as the angles $\a_1,\a_2$ in \eqref{alpha} are arbitrary.}, we define
\be\label{taunext}
\wh\tau_n(\l_1,...,\l_n;\nu):=\frac{\det \left[\sqrt{2\l_j}\e^{-2\l_j}\xi_k(\l_j)\right]_{j,k=1}^n}
{\Delta(\l_1,...,\l_n)}.
\ee

By construction $\wh \tau_n(\l_1,...,\l_n;\nu)$ has the same regular asymptotic expansion when the $\l_j$'s go to $\infty$ in every sector of the complex plane, see \eqref{asymptoticxi}. Notice that $\wh\tau_n(\l_1,...,\l_n;\nu)=\tau_n(\l_1,...,\l_n;\nu)$ provided that $\frac{\a_1}{2}<\arg\l_j<\frac{\a_2}{2}$.

\subsection{Schlesinger transformations}\label{secn}

Following the strategy already applied in \cite{BeCa2017,BeRu2017}, we consider a dressing of the bare ODE \eqref{bare}. This is conveniently expressed in terms of the  Riemann--Hilbert problem (RHP) \ref{problemn} below.

\noindent Fix $n\geq 0$, and $\l_1,...,\l_n\in\C\setminus\S$; from now on we imply dependence on this data. Introduce
\begin{align}
D_n(z)&:=\prod_{j=1}^n\begin{bmatrix}
\l_j+\sqrt{z} & 0 \\ 0 & \l_j-\sqrt{z}
\end{bmatrix}
\\
M_n(z)&:=D_n^{-1}(z_+) \e^{2\s_3\sqrt{z_+}} S(z) \e^{-2\s_3\sqrt{z_-}} D_n(z_-)
\end{align}
where the notation $\pm$ refers to the boundary values as in Fig. \ref{fig1}; the distinction between boundary values is only important along $z<0$. The matrices $M_n$ read more explicitly 
\be\label{Mn}
M_n(z)=\begin{cases}
\i\s_1 & z<0
\\
\begin{bmatrix}
1&0 \\ -\i\e^{\i\nu\pi}\e^{-4\sqrt{z}}\prod\limits_{j=1}^n\frac{\l_j+\sqrt{z}}{\l_j-\sqrt{z}} & 1
\end{bmatrix} & \arg z=\a_1
\\
\begin{bmatrix}
1&0 \\ -\i\e^{-\i\nu\pi}\e^{-4\sqrt{z}}\prod\limits_{j=1}^n\frac{\l_j+\sqrt{z}}{\l_j-\sqrt{z}} & 1
\end{bmatrix} & \arg z=\a_2.

\end{cases}
\ee
\noindent Notice that $M_n(z)=\1+\O\left(z^{-\infty}\right)$ as $z\to\infty$ along the rays $\arg z=\a_1,\a_2$.

\begin{shaded}
\begin{problem}\label{problemn}
Find a $2\times 2$ matrix $\G_n(z)=\G_n(z;\l_1,...,\l_n)$, analytic for $z\in\C\setminus\S$ satisfying the following jump condition along $\S$
\be\label{jumpn}
\G_n(z_+)=\G_n(z_-)M_n(z),
\ee
the growth condition at zero
\be\label{growthnzero}
\G_n(z)\sim \O(1)\Xi(z), \quad z\to 0,
\ee
and the normalization condition at infinity
\be\label{growthninfty}
\bg
\G_n(z)\sim z^{-\frac{\s_3}{4}}G Y_n(z) , \quad \ z\to\infty,
\\
G=\frac{1}{\sqrt{2}}\begin{bmatrix}
1 & -1 \\ 1& 1
\end{bmatrix}, \ \
 Y_n(z)=\1+\begin{bmatrix}
a_n & a_n \\ -a_n & -a_n
\end{bmatrix}
\frac{1}{\sqrt{z}}+\O\left(\frac{1}{z}\right)\in\GL\left(2,\C\left\llbracket\frac{1}{\sqrt{z}}\right\rrbracket\right) ,
\eg\ee
for some constant $a_n$ independent of $z$.
\end{problem}
\end{shaded}

\begin{remark}
The jump on the negative semi-axis $z<0$ in RHP \ref{problemn} is due to the multi--valuedness of $\sqrt{z}$. The position of this cut is completely arbitrary. By considering the analytic continuation beyond this cut we find that
\be
(z\e^{2\pi\i})^{-\frac{\s_3}{4}}GY_n(z\e^{2\pi\i})=z^{-\frac{\s_3}{4}}GY_n(z)\i\s_1
\ee
which in turn implies the following symmetry property
\be
\label{symmetry}
Y_n(z\e^{2\pi\i})=\s_1 Y_n(z)\s_1.
\ee
Hence the coefficients in front of even, resp. odd, powers of $\sqrt{z}$ have the form $\begin{bmatrix} u & v \\ v & u
\end{bmatrix}$, resp. $\begin{bmatrix} u& v \\-v& -u
\end{bmatrix}$.  
\end{remark}

\begin{remark}\label{remuniq}
The conditions \eqref{growthnzero} and \eqref{growthninfty} are required to ensure uniqueness of the solution to the RHP \eqref{problemn}. The growth condition \eqref{growthnzero} is necessary as the product of the jump matrices at $z=0$ is not the identity matrix. The necessity of the normalization condition \eqref{growthninfty} is explained as follows; indeed one may require the simpler boundary behaviour  $\G_n(z)\sim z^{-\frac{\s_3}{4}}G\left(\1+\mathcal{O}\left(z^{-1/2}\right)\right)$. However this would not uniquely fix the solution as follows from the identity
\be\label{identity}
\begin{bmatrix}
1 & 0\\ \beta & 1
\end{bmatrix}\ z^{-\frac{\s_3}{4}}\ G= z^{-\frac{\s_3}{4}}\ G\ \left(\1+\frac{1}{2}\begin{bmatrix}
\beta & -\beta \\ \beta &-\beta
\end{bmatrix} z^{-1/2}\right)
\ee
which would leave us with a one--parameter family of solutions, obtained one from the other by left multiplication by a matrix $\begin{bmatrix}
1 & 0\\ \beta & 1
\end{bmatrix}$, $\beta\in\C$. 
It follows from the same identity \eqref{identity} that the condition \eqref{growthninfty} removes this ambiguity.
This {\it gauge fixing} is chosen purely because of certain later convenience (see Lemma \ref{lemmapoly}) and  is otherwise entirely arbitrary. Indeed the tau function to be defined shortly (see Rem. \ref{remtau} below) is invariant under any transformation multiplying $\G_n$ on the left by an arbitrary constant (in $z$) matrix.
\end{remark}

The matrix $\Xi(z)\e^{-2\sqrt{z}\s_3}$ satisfies the jump condition \eqref{jumpn} and the growth condition \eqref{growthnzero} for $n=0$ but the asymptotic expansion \eqref{aaasymp} does not meet the requirement \eqref{growthninfty}. However we have
\be\label{expansion}
\begin{bmatrix}
 1 & 0 \\ \frac{3-8\nu+4\nu^2}{16} &1
\end{bmatrix}
\Xi(z)\e^{-2\sqrt{z}\s_3}\sim z^{-\frac{\s_3}{4}}G\left(\1+\frac{1-4\nu^2}{32\sqrt{z}}\begin{bmatrix}1 & 1\\ -1 & -1 \end{bmatrix}+\O\left(z^{-1}\right)\right), \ \ \ z\to\infty
\ee
which does fulfill \eqref{growthninfty}, with $a_0=\frac{1-4\nu^2}{32}$. Hence from now on we define
\be
\label{Gamma0}
\G_0(z):=\begin{bmatrix}
 1 & 0 \\ \frac{3-8\nu+4\nu^2}{16} &1
\end{bmatrix}
\Xi(z)\e^{-2\sqrt{z}\s_3}
\ee
which is by construction the solution to the full  RHP \ref{problemn} for $n=0$.

\noindent 
Suppose now that the solution $\G_n(z)$ to RHP \ref{problemn} exists; then the matrix
\be\label{Psin}
\Psi_n(z):=\G_n(z)D_n^{-1}(z)\e^{2\sqrt{z}\s_3}
\ee
has constant jumps along $\S$, therefore it satisfies a compatible system of linear ODEs
\be
\label{overdetn}
\Psi'_n(z)=A_n(z)\Psi_n(z), \ \ \frac{\pa \Psi_n(z)}{\pa \l_j}=\Omega_{j,n}(z)\Psi_n(z) \ \ (j=1,...,n)
\ee
where $A_n(z)$ is a rational function with simple poles at $z=0,\l_1^2,...,\l_n^2$ only while $\Omega_{1,n}(z)$,..., $\Omega_{n,n}(z)$ are rational functions with simple poles at $z=\l_1^2,...,\l_n^2$ only, as a consequence of the Liouville Theorem; compare with the growth condition \eqref{growthnzero}. The system \eqref{overdetn} is an isomonodromic system in the sense of \cite{JiMiUe1981}, whose tau function $\tau_n^I(\l_1,...,\l_n;\nu)$ \cite{JiMiUe1981,Be2010} is defined by
\begin{align}
\label{isomonodromicn}
\frac{\pa}{\pa \l_j}\log\tau_n^I(\l_1,...,\l_n;\nu)&=\frac{1}{2\pi\i}\int_\S\tr\left(\G_n^{-1}(z_-)\G'_n(z_-)\frac{\pa M_n(z)}{\pa \l_j}M_n^{-1}(z)\right)\d z
\\
&=\sum_{j=1}^n\res{z=\l_j}\tr\left(\G_n^{-1}(z)\G'_n(z)\frac{\pa D_n(z)}{\pa \l_j}D_n^{-1}(z)\right).
\nonumber
\end{align}

\begin{remark}
\label{remtau}
Notice that the expression \eqref{isomonodromicn} is not affected by a gauge transformation $\G_n(z)\to B \G_n(z)$, with $B\in\GL(2,\C)$ a $z$--independent nondegenerate matrix.
\end{remark}

\begin{shaded}
\begin{theorem}\label{thmmain}
We have
\be
\tau_n^I(\l_1,...,\l_n;\nu)=\wh \tau_n(\l_1,...,\l_n;\nu)
\ee
where $\tau^I_n(\l_1,...,\l_n;\nu)$ is defined in \eqref{isomonodromicn} and $\wh \tau_n(\l_1,...,\l_n;\nu)$ is defined in \eqref{taunext}.
\end{theorem}
\end{shaded}
\noindent The proof is contained in Sec. \ref{proofmain}.

In the terminology of \cite{JiMi1980}, the isomonodromic system \eqref{isomonodromicn} is obtained by a sequence of $n$ discrete Schlesinger transformations at the points $z=\l_1^2,...,\l_n^2$ of the ODE \eqref{bare}. We are applying here the RHP approach to Schlesinger transformations introduced in \cite{BeCa2015}.

\subsection{The limit $n\to\infty$}

Consider the (2,1)-entry of the jump matrix \eqref{Mn}; the following identity
\be
 \e^{-4\sqrt{z}}\prod\limits_{j=1}^n\frac{\l_j+\sqrt{z}}{\l_j-\sqrt{z}}
=
\exp\left(2\sum_{\ell\geq 0}\le[\left(\frac{1}{\l_1^{2\ell+1}}+\cdots+\frac{1}{\l_n^{2\ell+1}}\right) -2 \delta_{\ell,0}\ri]\frac{\sqrt{z}^{2\ell+1}}{2\ell+1}\right)
\ee
holds  uniformly over compact sets in $|z|< \min_j |\l_j|^2$. Together with the definition of the Miwa times \eqref{miwa} it suggests to consider the {\it phase function } 
\be
\vartheta(z;\t) := \sum_{\ell\geq 0}\left(t_\ell-2\delta_{\ell,0}\right)\sqrt{z}^{2\ell+1}. 
\ee
The Miwa times uniquely determine the $n$ values $\lambda_j$ up to permutations; however they clearly are not independent from each other for any fixed $n$, and therefore we want to explain in which sense we should understand the large--$n$ limit.  

Our main interest is in the computation of the higher--order log-derivatives of the gBGW tau function at $\t=0$ (Thm. \ref{thmcor}); however, the definition of analytic function of infinitely many variables is problematic, even more so for asymptotic expansions thereof. Therefore,  for our purposes it is sufficient to consider functions of only finitely many such variables by setting $t_{K+\ell}=0,  \ \ell\geq 0$,  for some $K$ sufficiently large, and then evaluating its log-derivatives. The ``inductive limit'' as $K\to\infty$ makes sense because ostensibly (as it will appear) the resulting formulas are  independent of $K$ as long as $K$ is large enough. 

It  is  clearly not possible to fix the value of infinitely many Miwa times given the $n$ values $\lambda_1,\dots,\lambda_n$, so the logic of an analytic proof should proceed as follows (see \cite{BeCa2017} for more details); we choose an appropriate sequence of matrices $\Lambda^{(n)}:= {\rm diag} (\lambda_1^{(n)} , \dots, \lambda_{n}^{(n)})$ such that the corresponding Miwa times $t_\ell(\Lambda^{(n)})$ tend, as $n\to \infty$, to a preassigned sequence ${\bf t}=(t_1,\dots, t_K, 0,0,0,0\dots)$. The fact that this is possible is a consequence of the Pad\'e\ approximation theorem for the function ${\rm e}^{\vartheta(z;{\bf t})}$. The limit of \eqref{taunext} as $n\to\infty$ is then considered as a function of finitely many Miwa times.  In this case it could be shown that it converges to the isomonodromic tau function for the RHP defined below (\ref{probleminfty})  in a suitable sector of the variables $(t_1, \dots, t_K)$. The computation of the limit of its log-derivatives at ${\bf t}=0$ (within the sector) results in the formulas of Thm. \ref{thmcor}, which are independent of the truncation parameter $K$; this is due ultimately to the formul\ae\ \eqref{isomonodromicinfty} and Lemma \ref{lemmanpoint} which express the log-derivatives of $\tau({\bf t})$ solely  in terms of the  solution $\Gamma(z; {\bf t}) $  of the RHP \ref{probleminfty}, together with the  fact that when the ($K$--truncated) ${\bf t}$ tends to zero within a  suitable sector, the solution $\Gamma(z;{\bf t})$ tends (uniformly) to the solution $\G_0(z)$ \eqref{Gamma0} of the  bare ODE. See also the last paragraph of this section. \vskip 3pt

The reader not interested in these analytical details, may consider the RHP \ref{probleminfty} directly as depending on infinitely many Miwa times and consider all subsequent manipulations as formal. 

Keeping this in mind we will dispose of these details and formally set 
\begin{align}
\vartheta(z;\t)&:=\sum_{\ell= 0}^K\left(t_\ell-2\delta_{\ell,0}\right)\sqrt{z}^{2\ell+1},
\\
M(z;\t)&:=\e^{-\vartheta(z_-;\t)\s_3}S(z)\e^{\vartheta(z_+;\t)\s_3}=
\begin{cases}
\i\s_1 & z<0
\\
\\
\begin{bmatrix}
1&0 \\ -\i\e^{\i\nu\pi}\e^{2\vartheta(z;\t)} & 1
\end{bmatrix} & \arg z=\a_1
\\
\\
\begin{bmatrix}
1&0 \\ -\i\e^{-\i\nu\pi}\e^{2\vartheta(z;\t)} & 1
\end{bmatrix} &\arg z=\a_2.
\end{cases}
\end{align}
We then consider the RHP \ref{probleminfty} below which is the (formal) reduction of RHP \ref{problemn} by setting to zero the Miwa times $t_{K+1}=t_{K+2}=\cdots=0$.

Therefore from now on we agree that $\t:=(t_0,t_1,...,t_K,0,0,...)$, where we remind that $K$ is fixed but arbitrary. We also assume that $t_K\not=0$ satisfies
\be\label{analyticcondition}
\Re\left(\sqrt{z}^{2K+1}t_K\right)<0, \text{ for }\arg z =\alpha_{1,2}
\ee
so that $M(z;\t)\sim \1+\O\left(z^{-\infty}\right)$ along $\arg z=\alpha_{1,2}$.

\begin{shaded}
\begin{problem}\label{probleminfty}
Find a $2\times 2$ matrix $\G(z;\t)$, analytic for $z\in\C\setminus\S$ satisfying the following jump condition along $\S$
\be\label{jumpinfty}
\G(z_+;\t)=\G(z_-;\t)M(z;\t),
\ee
the growth condition at zero
\be\label{growthinftyzero}
\G(z;\t)\sim \O(1)\Xi(z), \ \ z\to 0,
\ee
and the normalization condition at infinity
\be\label{growthinftyinfty}
\bg
\G(z;\t)\sim z^{-\frac{\s_3}{4}}G Y(z;\t) , \ \ \ z\to\infty,
\\
G=\frac{1}{\sqrt{2}}\begin{bmatrix}
1 & -1 \\ 1& 1
\end{bmatrix}, \ \
 Y(z;\t)=\1+\begin{bmatrix}
a(\t) & a(\t) \\ -a(\t) & -a(\t)
\end{bmatrix}
\frac{1}{\sqrt{z}}+\O\left(\frac{1}{z}\right)\in\GL\left(2,\C\left\llbracket\frac{1}{\sqrt{z}}\right\rrbracket\right) ,
\eg\ee
for some function $a(\t)$ of $\t$ independent of $z$.
\end{problem}
\end{shaded}

The considerations regarding the uniqueness exposed in Rem. \ref{remuniq} apply equally well here; the solution of RHP \ref{probleminfty} for $\t=(0,0,...)$ is $\G_0(z)$ defined in \eqref{Gamma0} by construction, satisfying \eqref{jumpinfty}, \eqref{growthinftyzero} and \eqref{growthinftyinfty} with $a(0,0,...)=\frac{1-4\nu^2}{32}$.

Repeating the arguments of Sec. \ref{secn}, assuming therefore that the unique solution $\G(z;\t)$ to RHP \eqref{probleminfty} exists, we get a compatible system of linear ODEs
\be
\label{overdetinfty}
\frac{\pa \Psi(z;\t)}{\pa z}=A(z;\t)\Psi(z;\t), \ \ \ \frac{\pa\Psi(z;\t)}{\pa t_\ell}=\Omega_\ell(z;\t)\Psi(z;\t), \ \ \ \ell=0,...,K
\ee
for the matrix
\be\label{Psiinfty}
\Psi(z;\t):=\G(z;\t)\e^{-\vartheta(z;\t)\s_3}.
\ee
More precisely we have the following Lemma, which is proven in Sec. \ref{secprooflimit}.

\begin{lemma}
\label{lemmapoly}
The matrices $\Omega_\ell(z;\t)$ are polynomials in $z$ of degree $\ell+1$ which can be written as
\be\label{Omega}
\Omega_\ell(z;\t)=-\left(\Psi(z;\t)\s_3\Psi^{-1}(z;\t)\sqrt{z}^{2\ell+1}\right)_+
\ee
where $()_+$ denotes the polynomial part\footnote{Note that by \eqref{symmetry} the expression $\Psi(z;\t)\s_3\Psi^{-1}(z;\t)\sqrt{z}^{2\ell+1}$ has an expansion in integer powers of $z$ only.} of a Laurent expansion in $z$ around $z=\infty$. The matrix $A(z;\t)$ is a rational matrix with a simple pole at $z=0$ which can be written as
\be\label{A}
A(z;\t)=\frac{1}{z}\left(-\frac{\s_3}{4}+\sum_{\ell\geq 0}\frac{2\ell+1}{2}(t_\ell-2\delta_{\ell,0})\Omega_\ell(z;\t)\right).
\ee
\end{lemma}

The system \eqref{overdetinfty} is again an isomonodromic system in the sense of \cite{JiMiUe1981} and its isomonodromic tau function $\tau^I(\t;\nu)$ is defined by
\begin{align}
\nonumber
\frac{\pa}{\pa t_\ell}\log\tau^I(\t;\nu)&=\frac{1}{2\pi\i}\int_\S\tr\left(\G^{-1}(z_-;\t)\G'(z_-;\t)\frac{\pa M(z;\t)}{\pa t_\ell}M^{-1}(z;\t)\right)\d z
\\
&=\res{z=\infty}\tr\left(\G^{-1}(z;\t)\G'(z;\t)\s_3\sqrt{z}^{2\ell+1}\right)\d z, \qquad \ell=1,...,K.
\label{isomonodromicinfty}
\end{align}
The meaning of the residue in \eqref{isomonodromicinfty} is formal and means simply (minus) the coefficient of the power $z^{-1}$ of a formal power series; in this regard we observe that $\G^{-1}(z;\t)\G'(z;\t)\s_3\sqrt{z}^{2\ell+1}$ is a power series in integer powers of $z$ only, thanks to \eqref{symmetry}.

Following  arguments similar to \cite[Prop. 3.6]{BeCa2017} we could also show that the solution of RHP \ref{probleminfty} exists in a domain of the form:  $|t_0|<2$, $\max_{j\geq 1} |t_j|<\epsilon$ (for some $\epsilon>0$) and $\arg t_{K}$ is a suitable range implied by \eqref{analyticcondition}. This would allow us to conclude that $\log \tau^I(\t;\nu)$ is analytic in the same domain (i.e. $\tau^I$ does not vanish)  and moreover that it admits an asymptotic expansion as $\t\to 0$ within the same domain. These considerations, while important, are not really necessary for the purposes of the present paper; in principle, the width of the domain of the asymptotic expansion indicates the Gevrey class of the function and hence the order of growth of the coefficients. 

In view of the above discussion we shall identify $\tau^I(\t;\nu)=\tau(\t;\nu)$ in all the formal computations below; in particular the proofs of Thm.s \ref{thmcor} and \ref{thmvir}, contained in Sec. \ref{secproofcor} and \ref{secproofvir} resp., exploit the expression for the logarithmic derivatives of the gBGW tau function in terms of the Jimbo--Miwa--Ueno formula, i.e. of the second line in \eqref{isomonodromicinfty}.

\subsection{KdV and Painlev\'{e} XXXIV hierarchies}

It is well known that the Kontsevich--Witten KdV tau function \cite{Wi1991,Ko1992} provides a solution to the Painlev\'{e} I hierarchy \cite{DoSh1990,BeCa2017}. Here we observe that the gBGW tau function provides in the same way a solution to the Painlev\'{e} XXXIV hierarchy.

More precisely,  let us call $x:=t_0$ and introduce
\be\label{defu}
u(x,\t_{\geq 1};\nu):=\frac{\pa^2}{\pa x^2}\log\tau(x,\t_{\geq 1};\nu), \ \ \ \t_{\geq 1}:=(t_1,t_2,...)
\ee
which is a solution to the KdV hierarchy
\be\label{kdv}
\frac{\pa u}{\pa t_{\ell}}=\frac{\d}{\d x}\mathcal{L}_{\ell+1}[u], \ \ \ \ell\geq 1
\ee
satisfying the initial condition
\be
u(x,\t_{\geq 1}=0;\nu)=\frac{1-4\nu^2}{8(2-x)^2}
\ee
as we shall compute below in Sect. \ref{reminitial}, see \eqref{initial}. In \eqref{kdv} we denote $\mathcal{L}_\ell[u]$ the Lenard--Magri differential polynomials, normalized as
\be
\mathcal{L}_0[u]=1, \ \ \begin{cases}\frac{\d}{\d x}\mathcal{L}_{\ell+1}=\left(\frac{1}{4}\frac{\d^3}{\d x^3}+2u\frac{\d}{\d x}+u_x\right)\mathcal{L}_\ell[u] \\ 
\mathcal{L}_{\ell+1}[u=0]=0\end{cases}\text{ for }\ell\geq 0.
\ee
Let us now write the Virasoro constraint $L_0\tau=0$, see \eqref{vir}, as
\be
(x-2)\frac{\pa\log\tau}{\pa x}+\sum_{\ell\geq 1}(2\ell+1)t_{\ell}\frac{\pa \log\tau}{\pa t_{\ell}}+\frac{1-4\nu^2}{8}=0
\ee
and taking two derivatives in $x$ we have
\be
(x-2)\frac{\pa^3\log\tau}{\pa x^3}+2\frac{\pa^2\log\tau}{\pa x^2}+\sum_{\ell\geq 1}(2\ell+1)t_{\ell}\frac{\pa^3 \log\tau}{\pa_x^2 \pa t_{\ell}}=0.
\ee
The following proposition then follows from the definition \eqref{defu} of $u$ and the KdV hierarchy equations \eqref{kdv}.

\begin{shaded}
\begin{proposition}\label{proppain}
If we set $t_{\ell}=0$ for $\ell\geq K+1$ as above, then $u(x;t_1,...,t_K,0,...;\nu)$ solves the $K$th member of the PXXXIV hierarchy;
\be\label{painleve}
2u+(x-2)u_x+\sum_{\ell=1}^K(2\ell+1)t_\ell \frac{\d}{\d x}\mathcal{L}_{\ell+1}[u]=0
\ee
which is an ODE in $x$, where $t_1,...,t_K$ are regarded as parameters.
\end{proposition}
\end{shaded}

The Painlev\'{e} XXXIV hierarchy has been considered in \cite{ClJoPi1999} and it is related by a Miura transformation to the Painlev\'{e} II hierarchy, first introduced in \cite{FlNe1980}.

For example, the case $K=1$ in \eqref{painleve} is
\be\label{u}
\frac 3 4 t_1u_{xxx}+9t_1uu_x+(x-2)u_x+2u=0.
\ee
By the simple scaling
\be
x=2-\left(\frac{3t_1}{4}\right)^{\frac 1 3}y, \qquad u(x)=\left(\frac{2}{9t_1^2}\right)^{\frac 1 3}v(y)
\ee
\eqref{u} reads
\be\label{v}
v_{yyy}+6vv_y-yv_y-2v=0
\ee
which we call, following the literature, see e.g. \cite{ClJoPi1999}, the Painlev\'{e} XXXIV equation. 

It is known \cite{In1956,AbFo1982} that \eqref{v} is equivalent to the Painlev\'{e} II equation
\be\label{pii}
w_{yy}=w^3+y w+\alpha,
\ee
in the sense that 
the Miura transformation
\be
v=-w^2-w_y, \ \ \ w=\frac{v_y+\alpha}{2v-y}
\ee
is  a one--to--one map between solutions to \eqref{v} and to \eqref{pii}.
  
Using \eqref{expansionY}, \eqref{Omega0} and \eqref{Omega1} we can write down explicitly the Lax pair for \eqref{u} as
\be
A=A_1z+A_0+\frac{A_{-1}}{z}, \ \ \ \Omega=\begin{bmatrix}
-2a & -1\\-z-2a_x+4a^2 & 2a
\end{bmatrix}
\ee
where
\begin{align}
A_1&=\begin{bmatrix}0 & 0 \\ -\frac{3t_1}{2} & 0
\end{bmatrix}, \qquad
A_0=
\begin{bmatrix}
 -3 t_1 a & -\frac{3t_1}{2}\\
 6 t_1 a^2+3 t_1 a_x-\frac{x}{2}+1 & 3 t_1 a 
\end{bmatrix}, \\
\nonumber A_{-1}&=\left[\begin{smallmatrix}
 -(x-2) a-6 t_1 a_x a-\frac{3}{2} t_1 a_{xx}-\frac{1}{4} & -\frac{x-2}{2}-3 t_1 a_x \\
 2 (x-2) a^2+12 t_1 a_x a^2+6 t a_{xx} a+a+12 t_1 a_x^2+2 (x-2) a_x+\frac{3}{2} t_1 a_{xxx} & (x-2) a+6 t_1 a_x a+\frac{3}{2} t_1 a_{xx}+\frac{1}{4} 
\end{smallmatrix}\right].
\end{align}
Indeed, the compatibility of $\Psi'=A\Psi$ and $\Psi_x=\Omega\Psi$ implies the zero curvature condition
\be
A_x - \Omega' -[\Omega,A]=\frac 1 z \begin{bmatrix}
 0 & 0 \\
\frac{3}{2} t_1 a_{xxxx}+36 t_1 a_{xx} a_x+2 (x-2) a_{xx}+4 a_x & 0 \\
\end{bmatrix}=0
\ee
which, identifying $u=2a_x$ from \eqref{pa0}, gives \eqref{u}. Setting $t_1=-\frac{4}{3}$, $x-2=y$ and $4a(x)=\alpha(y)$ we obtain the following Lax pair for \eqref{v};
\begin{align}
\label{laxpair}
A&=\renewcommand*{\arraystretch}{1.25}\begin{bmatrix}
\a+\frac{2\a_{yy}-y\a+2\a\a_y-1}{4z}
&
2+\frac{2\a_y-y}{2z}
\\
2z-\frac{y}{2}-\frac{\a^2}{2}-\a_y+\frac{2\a+y\a^2+4y\a_y-2\a^2\a_y-8\a_y^2-4\a\a_{yy}-4\a_{yyy}}{8z}
& 
-\a-\frac{2\a_{yy}-y\a+2\a\a_y-1}{4z} 
\end{bmatrix},\nonumber
\\
\Omega&=\renewcommand*{\arraystretch}{1.25}\begin{bmatrix}
-\frac{\a}{2} & -1 \\ -z+\frac{\a^2}{4}+\frac{\a_y }{2} & \frac \a 2
\end{bmatrix},\qquad \begin{cases}\Psi'=A\Psi \\  \Psi_y=\Omega\Psi \end{cases} \Rightarrow \ \ \a_{yyyy}+6\a_y\a_{yy}-y\a_{yy}-2\a_y=0
\end{align}
which is \eqref{v} for $v:=\a_y$. Finally we note that after a gauge transformation on \eqref{laxpair} of the form $\wh A=GAG^{-1}$, $\wh\Omega=G\Omega G^{-1}+G_yG^{-1}$ with $G=\begin{bmatrix}
1 & 0 \\ \frac \a 2 & 1
\end{bmatrix}$ we obtain a Lax pair
\be
\wh A=
\begin{bmatrix}
 \frac{2v_y-1}{4z} & \frac{2v-y}{2z}+2  \\
2 z-v-\frac{y}{2}+\frac{-2 v^2+y v-v_{yy}}{2 z} & \frac{1-2v_y}{4z} \\
\end{bmatrix},
\ \ \ \
\wh \Omega=\begin{bmatrix}0 & -1 \\ v-z & 0 \end{bmatrix}
\ee
for \eqref{v} in $v$ directly. 

\subsubsection{The bare tau function} \label{reminitial}
We now compute the ``bare'' tau function for $\t=(x,0,0,\dots)$ using the solution of the bare RHP. 
The time $x=t_0$ is related to scalings of the variable $z$ in the RHP \ref{probleminfty}; hence, restricting to real values of $x$ for simplicity, we have 
\be
\G(z;(x,0,...))=\left(1-\frac{x}{2}\right)^{\frac{\s_3}{2}}\G_0\left(\left(1-\frac{x}{2}\right)^2z\right)
\ee
where we assume $-2<x<2$ and take the principal branch of the square roots. At the level of asymptotic expansions, we are replacing $\sqrt{z}\to\left(1-\frac{x}{2}\right)\sqrt{z}$ in the asymptotic expansion of $\G_0(z)$; from \eqref{expansion} we see that
\be\label{last}
\G(z;(x,0,...))\sim z^{-\frac{\s_3}{4}}G\left(\1+\frac{1-4\nu^2}{32\left(1-\frac{x}{2}\right)\sqrt{z}}\begin{bmatrix} 1 & 1 \\ -1 & -1
\end{bmatrix}
+\O\left(z^{-1}\right)\right), \ \ \ z\to \infty.
\ee
Using \eqref{last} a direct computation shows that
\be\label{taux}
\pa_x\log\tau(x,0,...)=\res{z=\infty}\tr\left(\G^{-1}(z;(x,0,...))\G'(z;(x,0,...))\s_3\sqrt{z}\right)\d z=\frac{1-4\nu^2}{8(2-x)}
\ee
which provides the initial datum for the KdV hierarchy \eqref{kdv};
\be\label{initial}
u(x;\t_{\geq 1}=0;\nu)=\frac{\pa^2}{\pa x^2}\log\tau(x,0,...)=\frac{1-4\nu^2}{8(2-x)^2}.
\ee
Moreover \eqref{taux} implies that
\be
\tau(x,0,...)=C(2-x)^{\frac{4\nu^2-1}{8}}
\ee
for some nonvanishing integration constant $C\not=0$, which indicates that RHP \ref{probleminfty} for $\t=(x,0,...)$ is solvable for all values of $x\neq 2$.

\paragraph*{Acknowledgements.} 
The work of M.B. is supported in part by the Natural Sciences and Engineering Research Council of Canada (NSERC) grant RGPIN-2016-06660 and  by the   FQRNT grant ``Applications des syst\`emes int\'egrables \`a\ les surfaces de Riemann et les espaces de modules'' (2016-PR-190918).
G.R. wishes to thank the Department of Mathematics and Statistics at Concordia University where this work was carried out. This project has received funding from the European Union's H2020 research and innovation programme under the Marie Sklowdoska-Curie grant No. 778010 {\em  IPaDEGAN}.

\paragraph*{Note.} 
During the submission phase we were made aware that some of the formul\ae\ (Thm. \ref{thmcor}) will appear in a forthcoming work by B. Dubrovin, D. Yang and  D. Zagier \cite{toappear}. The methods employed in the respective papers are however substantially different.

\section{Proofs}\label{secproof}

\subsection{Proof of Thm. \ref{thmmain}}\label{proofmain}

In this section we prove Thm. \ref{thmmain}; the approach is exactly parallel to that in \cite[App. A]{BeCa2017}, which we refer to for further details (see also \cite{BeRu2017}).

\subsubsection{The characteristic matrix}

Following \cite{BeCa2015} we introduce the characteristic matrix
\be
\label{charmatr}
\mathcal{G}=[\mathcal{G}_{j,k}]_{j,k=1}^n, \ \ \
\mathcal{G}_{j,k}=\begin{cases}
-\res{z=\infty}\frac{z^k}{z-\l_j^2}\mathbf{e}_2^\top\G_0^{-1}(\l_j^2)\G_0(z)G^{-1}z^{\frac{\s_3}{4}}\mathbf{e}_{1+k}
&\text{if }-\frac\pi 2<\arg\l<\frac\pi 2
 \\
-\res{z=\infty}\frac{z^k}{z-\l_j^2}\mathbf{e}_1^\top\G_0^{-1}(\l_j^2\e^{\mp 2\pi\i})\G_0(z)G^{-1}z^{\frac{\s_3}{4}}\mathbf{e}_{1+k}
&\text{if }\frac\pi 2 <\pm\arg\l<\pi \end{cases}
\ee
where $\mathbf{e}_1=\begin{bmatrix}
1 \\ 0
\end{bmatrix}$, $\mathbf{e}_2=\begin{bmatrix}
0\\ 1
\end{bmatrix}$, and the index in $\mathbf{e}_{1+k}$ is understood $\mod 2$ (e.g. $\mathbf{e}_3=\mathbf{e}_1$, $\mathbf{e}_4=\mathbf{e}_2$); $\G_0(z)$ is as in \eqref{Gamma0}, and note that the gauge factor of \eqref{Gamma0} is irrelevant here, as $\mathcal{G}_{j,k}$ is invariant under $\G_0\mapsto B\G_0$ for any $B\in\GL(2,\C)$.

The residue in \eqref{charmatr} is by definition a formal residue, i.e. we regard
\be\label{formalpowerseries}
\G_0(z)G^{-1}z^{\frac{\s_3}{4}}=z^{-\frac{\s_3}{4}}GY_n(z)G^{-1}z^{\frac{\s_3}{4}}=\1+\O\left(z^{-1}\right)\in\GL\left(2,\C\left\llbracket z^{-1}\right\rrbracket\right)
\ee
as a formal power series and the formal residue is simply the coefficient of $z^{-1}$. It can be checked that thanks to the property \eqref{symmetry} the expression \eqref{formalpowerseries} contains integer powers of $z$ only.

\begin{proposition}
We have
\be\label{det}
\det \mathcal{G}=C \det\left[\e^{-2\l_j}\xi_k(\l_j)\right]_{j,k=1}^{n}
\ee
where the proportionality constant $C$ (irrelevant in the following) is
\be
C:=(-1)^{\left \lfloor\frac a 2 \right\rfloor}(-\i)^{b_+}\i^{b_-},\ \ \
a:=\sharp\left\lbrace j: \ -\frac{\pi}{2}\arg\l_j<\frac{\pi}{2}\right\rbrace,\ \ b_\pm:=\sharp\left\lbrace j: \ \frac{\pi}{2}<\pm\arg\l_j<\pi\right\rbrace.
\ee
\end{proposition}

\noindent {\bf Proof.} 
Let us consider the case $-\frac\pi 2<\arg\l_j<\frac\pi 2$ first; by the definition \eqref{charmatr} and simple algebra using \eqref{eqxi}, we see that the $(2m+1)$th, resp. $(2m+2)$th, column of $\mathcal{G}$ is the second, resp. the first, entry in the row vector coefficient of $z^{-m}$ in
\be\label{veramenteboh}
\frac{\e^{-2\l_j}}{1-\frac{\l_j^2}{z}}[-\xi_2(\l_j),\xi_1(\l_j)]\left(\1+\O\left(z^{-1}\right)\right)=\sum_{m\geq 0}\frac{\e^{-2\l_j}\l_j^{2m}}{z^m}[-\xi_2(\l_j),\xi_1(\l_j)]\left(\1+\O\left(z^{-1}\right)\right),
\ee
where $j$ is the row index of the columns of $\mathcal{G}$. Hence we note that the first column of $\mathcal{G}$ is given by $[\e^{-2\l_j}\xi_1(\l_j)]_{j=1}^n$ and the second one by $[-\e^{-2\l_j}\xi_2(\l_j)]_{j=1}^n$. 

For the next columns we proceed by induction. Indeed, as the $\mathcal{O}(z^{-1})$ term in \eqref{veramenteboh} does not depend on the row index $j$, it follows that the $(2m+1)$th column is $[\e^{-2\l_j}\l^{2m}_j\xi_1(\l_j)]_{j=1}^n$ up to a linear combination of the previous (odd) column. Similarly the $(2m+2)$th column is $[-\e^{-2\l_j}\l_j^{2m}\xi_2(\l_j)]_{j=1}^n$ up to a linear combination of the previous (even) columns. Now we recall \cite{AbSt1965}
\be\label{recursionik}
I_{\a+1}(2\l)=I_{\a-1}(2\l)-\frac{\a}{\l}I_{\a}(2\l)
,\ \ \ 
K_{\a+1}(2\l)=K_{\a-1}(2\l)+\frac{\a}{\l}K_{\a}(2\l)
\ee
which implies
\be\label{recursionxi}
\xi_{k+2}(\l)=\l^2\xi_{k}(\l)-(k-\nu)\xi_{k+1}(\l)\text{ when }-\frac\pi 2<\arg\l<\frac\pi 2
\ee
and so
\begin{align}
\l^{2m}\xi_{1}(\l)&\equiv \xi_{2m+1}(\l)\ \mod \left(\xi_1(\l),...,\xi_{2m}(\l)\right) \nonumber
\\
\label{rec}
\l^{2m}\xi_2(\l)&\equiv \xi_{2m+2}(\l) \ \mod \left(\xi_1(\l),...,\xi_{2m+1}(\l)\right) \quad (m\geq 1).
\end{align}
It follows that the matrices $\mathcal{G}$ and $[(-1)^{k-1}\e^{-2\l_j}\xi_k(\l_j)]_{j,k=1}^m$ differ by multiplication by a unimodular matrix, more precisely by a triangular matrix with $1$'s along the diagonal; in particular they have the same determinant and Proposition is proven when $-\frac\pi 2<\arg\l_j<\frac\pi 2$.

The case when $\frac\pi 2 < \pm\arg\l_j<\pi$ is completely analogous so we just briefly comment on the differences; expression \eqref{veramenteboh}, in view of \eqref{charmatr} and \eqref{eqxi}, must be replaced by
\be
\frac{\e^{-2\l_j}}{1-\frac{\l_j^2}{z}}[\pm\i\xi_2(\l_j),\pm\i\xi_1(\l_j)]\left(\1+\O\left(z^{-1}\right)\right)=\sum_{m\geq 0}\frac{\e^{-2\l_j}\l_j^{2m}}{z^m}[\pm\i\xi_2(\l_j),\pm\i\xi_1(\l_j)]\left(\1+\O\left(z^{-1}\right)\right)
\ee
while the recursion \eqref{recursionxi} must be replaced by
\be
\xi_{k+1}(\l)=\l^2\xi_{k-1}(\l)+(k-\nu)\xi_k(\l),\ \ \ \ \frac\pi 2<\pm\arg\l<\pi
\ee
which is again a consequence of \eqref{recursionik}. Hence \eqref{rec} holds true in the case $\frac\pi 2 < \pm\arg\l_j<\pi$ as well and as above, taking care of the $\pm$'s and $\pm\i$'s, we have the thesis. \QED

\subsubsection{Schlesinger transform and Malgrange form}

\begin{proposition}Suppose RHP \ref{problemn} has a solution $\G_n(z)$. Then there exists a rational matrix $R_n(z)$ with simple poles at $z=\l_1^2,...,\l_n^2$ only such that
\be
\G_n(z)=R_n(z)\G_0(z)D_n(z).
\ee
\end{proposition}

\noindent {\bf Proof.} 
It can be checked that $R_n(z):=\G_n(z)D^{-1}_n(z)\G_0^{-1}(z)$ does not have jumps along $\S$, while having at worse simple poles at $z=\l_1^2,...,\l_n^2$; the thesis is now a consequence of Lioville's Theorem.
\QED

Hereafter we employ the short notation $\pa_j:=\frac{\pa}{\pa \l_j}$ and we consider the case $\Re\l_j\geq 0$ only for clarity's sake; the general case is a straightforward generalization.

The following variational formula has been proven in \cite[App. B]{BeCa2015};
\be\label{variational}
\bg
\pa_j\log\det \mathcal{G}=\sum_{k=1}^n\res{z=\l_k^2}\tr\left(R^{-1}_nR'_n\pa_j J_kJ_k^{-1}\right)+\res{z=\infty}\tr\left(R^{-1}_nR'_n\pa_j J_\infty J_\infty^{-1}\right)+
\\
+\sum_{k=1}^n\res{z=\l_k^2}\tr\left(\G_0^{-1}\G_0'\pa_j U_k U_k^{-1}\right)
\eg
\ee
where
\be
\bg
J_k:=\G_0(z)\begin{bmatrix}1&0 \\ 0& \l_k^2-z\end{bmatrix}, \ \ \
J_\infty:=\G_0(z)D_n(z)G^{-1}z^{\frac{\s_3}{4}}, \ \ \
U_k:=\begin{bmatrix}1 & 0\\ 0& z-\l_k^2\end{bmatrix}, \ \ \ \ k=1,...,n.
\eg
\ee

We are ready to give the proof of Thm. \ref{thmmain}; let us compute the Malgrange form
\be
\omega_n(\pa_j):=\frac{1}{2\pi\i}\int_\S \tr\left( \G_n(z_-)^{-1} \G_n'(z_-)\pa_j M_n(z)M_n^{-1}(z)\right) \d z
\ee
by using $\G_n=R_n\G_0 D_n$ and $M_n=D_n^{-1}M_0D_n$ where $M_0(z):=\e^{2\sqrt{z_-}\s_3}S(z)\e^{-2\sqrt{z_+}\s_3}$. After some elementary steps\footnote{Which are explained in detail in \cite{BeCa2015,BeCa2017,BeRu2017}.} we obtain
\be
\omega_n(\pa_j)=\sum_{z_*\in\left\lbrace\l_1^2,...,\l_n^2,\infty\right\rbrace}\res{z=z_*}\tr\left(R_n^{-1}R_n'\G_0\pa_j D_nD^{-1}_n\G_0^{-1}+\G_0^{-1}\G_0'\pa_j D_nD^{-1}_n\right)
\ee
and by using the identities
\be
\pa_j J_\infty J_\infty^{-1}=\G_0\pa_j D D^{-1}\G_0^{-1}
\ee
we obtain (comparing with \eqref{variational})
\be\label{boooh}
\omega_n(\pa_j)=\pa_j\log\det\mathcal{G}+\sum_{k=1}^n\res{z=\l_k^2}\tr\left(\G_0^{-1}R_n^{-1}R_n'\G_0(\pa_j D_n D_n^{-1}-\pa_j U_k U_k^{-1})\right)
\ee
as $\res{z=\infty}\tr\left(\G_0^{-1}\G_0'\pa_jD_n D_n\right)=0$. Introducing now the matrices
\be
T_k:=D_nU^{-1}_k=\begin{bmatrix}1&0 \\ 0&\frac{\prod_{k'\not=k}(\l_{k'}-\sqrt{z})}{\l_k+\sqrt{z}}\end{bmatrix}, \ \ R_k^+:=R_n\G_0U_k, \ \ \ k=1,...,n
\ee
which are analytic at $z=\l_k^2$ and satisfy $\pa_ jD_nD_n^{-1}-\pa_j U_kU_k^{-1}=\pa_jT_kT_k^{-1}$ we compute each summand in the right--hand side of \eqref{boooh} as
\begin{align}
&\res{z=\l_k^2}\tr\left(R^{-1}_nR'_n\G_0\pa_j T_k T^{-1}_k\G^{-1}_0\right)=\res{z=\l_k^2}\tr\left((U_k^{-1}\G^{-1}_0R_n^{-1})(R'_n\G_0 U_k)\pa_j T_k T^{-1}_k\right) \nonumber
\\
&\qquad =\res{z=\l_k^2}\tr\left((U_k^{-1}\G^{-1}_0R^{-1}_n)((R_n\G_0 U_k)'-R_n\G'_0 U_k-R_n\G_0 U_k')\pa_j T_k T^{-1}_k\right)\nonumber
\\
&\qquad =\underbrace{\res{z=\l_k^2}\tr\left((R_k^+)^{-1}(R^+_k)'\pa_j T_k T^{-1}_k\right)}_{=0}-
\underbrace{\res{z=\l_k^2}\tr\left(\G^{-1}_0\G'_0\delta T_k T^{-1}_k\right)}_{=0}-
\res{z=\l_k^2}\tr\le(U_k^{-1}U'_k\pa_j T_k T^{-1}_k\right)\nonumber
\\
&\qquad=-\res{z=\l_k^2}\frac{1}{z-\l_k^2}\left(\pa_j\left(\frac{\prod_{k'\not=k}(\l_{k'}-\sqrt{z})}{\l_k+\sqrt{z}}\right)\frac{\l_k+\sqrt{z}}{\prod_{k'\not=k}(\l_{k'}-\sqrt{z})}\right)\nonumber
\\
&\qquad=-\res{z=\l_k^2}\frac{1}{z-\l_k^2}\times\begin{cases}
\frac{1}{\l_j-\sqrt{z}} & \text{if }j\not=k\\ \frac{1}{\l_j-\sqrt{z}} &\text{if }j=k 
\end{cases}
=\begin{cases}
\frac{1}{\l_k-\l_j} & \text{if }j\not=k\\ \frac{1}{2\l_k} &\text{if }j=k.
\end{cases}
\end{align}
From \eqref{boooh} we get, after a simple integration,
\be
\omega_n(\pa_j)=\pa_j\log\left(\frac{\prod_{j=1}^n\sqrt{\l_j}}{\Delta(\l_1,...,\l_n)}\det\mathcal{G}\right).
\ee
In view of \eqref{det} and \eqref{taunext} the proof of Thm. \ref{thmmain} is complete by observing that the isomonodromic tau function is defined only up to multiplicative constants by $\pa_j\log\tau_n^I=\omega(\pa_j)$, see \eqref{isomonodromicn}.

\subsection{Proof of Lemma \ref{lemmapoly}}\label{secprooflimit}

In this proof we omit the dependence on $(z;\t)$. The matrix $\Omega_\ell=\frac{\pa\Psi}{\pa t_\ell}\Psi^{-1}$ (with $\Psi$ as in \eqref{Psiinfty}) has no jumps along $\S$. In principle it may have an isolated singularity at $z=0$ (a pole or worse); however  this cannot happen because of condition \eqref{growthinftyzero}. Therefore $\Omega_\ell$ has a removable singularity at $z=0$ and thus extends to an entire function. 
From inspection of the asymptotic behaviour of $\Psi$  at $\infty$,  it follows that $\Omega_\ell$ is an entire function of $z$ with polynomial growth at $z=\infty$. By the Liouville Theorem $\Omega_\ell$ is a polynomial of $z$, which coincides then with the polynomial part of its asymptotic expansion;
\be\bg
\Omega_\ell=\left(\frac{\pa\Psi}{\pa t_\ell}\Psi^{-1}\right)_+
=\underbrace{\left(z^{-\frac{\s_3}{4}}G\frac{\pa Y}{\pa t_\ell}Y^{-1}G^{-1}z^{\frac{ \s_3}{ 4}}\right)_+}_{=0}-\left(\Psi\s_3\Psi^{-1}\frac{\pa \vartheta}{\pa t_\ell}\right)_+=-\left(\Psi\s_3\Psi^{-1}\sqrt{z}^{2\ell+1}\right)_+
\eg\ee
where the first term vanishes thank to our choice of normalization in \eqref{growthinftyinfty}.

The same reasoning applies to $A=\Psi'\Psi^{-1}$, with the only exception that, in view of growth condition at $z=0$ \eqref{growthinftyzero}, $A$ has a simple pole at $z=0$. It follows by the Liouville Theorem that $A$ is a rational function of $z$, which coincides then with the Laurent expansion at $\infty$ truncated at the term in $z^{-1}$; namely
\begin{align}
A&=\frac{1}{z}\left(z\Psi'\Psi^{-1}\right)_+
=-\frac{\s_3}{4z}+\underbrace{\frac{1}{z}\left(z z^{-\frac{\s_3}{4}}GY'Y^{-1}G^{-1}z^{\frac{\s_3}{4}}\right)_+}_{=0}-\frac{1}{z}\left(z\Psi\s_3\Psi^{-1}\vartheta'\right)_+
\\
&=-\frac{\s_3}{4z}-\sum_{\ell\geq 0}\frac{2\ell+1}{2z}(t_\ell-2\delta_{\ell,0})\left(z\Psi\s_3\Psi^{-1}\sqrt{z}^{2\ell-1}\right)_+
=\frac{1}{z}\left(-\frac{\s_3}{4}+\sum_{\ell\geq 0}\frac{2\ell+1}{2}(t_\ell-2\delta_{\ell,0})\Omega_\ell\right)
\nonumber
\end{align}
where again the term indicated vanishes thank to our choice of normalization in \eqref{growthinftyinfty}.

\begin{remark}
The expression \eqref{A} for $\t=0$ coincides with the ODE \eqref{bare} up to the gauge transformation \eqref{Gamma0}; indeed, using the expression \eqref{Omega0} below for $\Omega_0$ and the initial conditions $a(0,0,...)=\frac{1-4\nu^2}{32}$, $c(0,0,...)=-\frac{(9-4\nu^2)(1-4\nu^2)}{512}$ (which are read off the expansion of $\G_0(z)$) we see that \eqref{A} reduces to
\be\renewcommand*{\arraystretch}{1.25}
A(z)=-\frac{\s_3}{4z}-\frac{\Omega_0}{z}=
\begin{bmatrix}-\frac{3+4\nu^2}{16z} & \frac 1 z \\ 1-\frac{9-40\nu^2+16\nu^4}{256z} & \frac{3+4\nu^2}{16z}
\end{bmatrix}=\begin{bmatrix} 1 & 0\\ \frac{3-8\nu+4\nu^2}{16} & 1 \end{bmatrix}
\begin{bmatrix}
-\frac{\nu}{2z} & \frac 1 z  \\ 1 & \frac{\nu}{2z}
\end{bmatrix} 
\begin{bmatrix}
1 & 0 \\ -\frac{3-8\nu+4\nu^2}{16} & 1
\end{bmatrix}.
\ee
\end{remark}

\subsection{Proof of Thm. \ref{thmcor}}\label{secproofcor}

The proof of Thm. \ref{thmcor} follows from the same algebraic manipulations first introduced in \cite{BeDuYa2016} which have subsequently appeared many times, e.g. in \cite{BeDuYa2017,DuYa2017,BeRu2017} and it is explained in detail for the reader's convenience.

\subsubsection{One--point function}

We use \eqref{isomonodromicinfty} to compute
\begin{align}\label{s1t}
\sum_{\ell\geq 0}z^{-\ell-1}\frac{\pa\log\tau(\t)}{\pa t_\ell}&=\sum_{\ell\geq 0}z^{-1-\ell}\res{w=\infty}\tr\left(\G^{-1}(w;\t)\G'(w;\t)\s_3\sqrt{w}\right)w^\ell\d w
\\
&=-\tr\left(\G^{-1}(z;\t)\G'(z;\t)\s_3\sqrt{z}\right)
\\
&=-\tr\left(\sqrt{z}\Psi^{-1}(z;\t)\Psi'(z;\t)\s_3\right)-2\sqrt{z}\vartheta'(z;\t)
\nonumber\end{align}
where we have used $\G=\Psi\e^{\vartheta\s_3}$; evaluation at $\t=0$ of \eqref{s1t} gives, recalling definition \eqref{defsn},
\be
\label{s1}
S_1(z;\nu)=2-\tr\left(\sqrt{z}\,\Xi^{-1}(z)\Xi'(z)\s_3\right)=2-\tr\left(\sqrt{z}\ \begin{bmatrix} -\frac{\nu}{2z} & \frac 1 z \\ 1 & \frac{\nu}{2z}
\end{bmatrix}\Xi(z)\s_3\Xi^{-1}(z)\right)
\ee
where $\Xi(z)$ has been defined in \eqref{Psi0}, and we have used the ODE \eqref{bare}; in \eqref{s1} we identify $\Xi(z)$ with its asymptotic expansion at $z=\infty$.

\begin{lemma}\label{lemmaU}
We have, at the level of asymptotic expansions,
\be
\sqrt{z}\,\Xi(z)\s_3\Xi^{-1}(z)=\mathcal{U}(z;\nu)
\ee
where $\mathcal{U}(z;\nu)$ is defined in \eqref{U}.
\end{lemma}

\noindent {\bf Proof.} We compute $\mathcal{U}(z;\nu)$ in the sector $\a_1<\arg z<\a_2$, the result holds in every sector due to the fact that $\Xi(z)$ has the same asymptotic expansion in every sector by construction. Hence we compute
\be
\sqrt{z}\,\Xi\s_3\Xi^{-1}=\sqrt{z}\begin{bmatrix}
\mathcal{U}_{11} & \mathcal{U}_{12} \\
\mathcal{U}_{21} & -\mathcal{U}_{11}
\end{bmatrix}
,\quad \begin{cases}
\mathcal{U}_{11}:=2\sqrt{z}\left(I_{-\nu}(2\sqrt{z})K_{1-\nu}(2\sqrt{z})-I_{1-\nu}(2\sqrt{z})K_{-\nu}(2\sqrt{z})\right) \\
\mathcal{U}_{12}(z):=4I_{-\nu}(2\sqrt{z})K_{-\nu}(2\sqrt{z})\\
\mathcal{U}_{21}:=4zI_{1-\nu}(2\sqrt{z})K_{1-\nu}(2\sqrt{z}). 
\end{cases}\ee
From the ODE \eqref{bare} we deduce
\be
\left(\frac{\U}{\sqrt{z}}\right)'=\left[A,\frac{\U}{\sqrt{z}}\right], \quad \ A=\begin{bmatrix}
-\frac{\nu}{2z} & \frac 1 z \\ 1 & \frac{\nu}{2z}
\end{bmatrix}
\ee
from which we obtain the system of ODEs
\be\label{system}
\begin{cases}
2z\U_{11}'=-2z\U_{12}+2\U_{21}
\\
2z\U_{12}'=-4\U_{11}-2\nu\U_{12}
\\
2z\U_{21}'=4z\U_{11}+2\nu\U_{21}.
\end{cases}
\ee
Consider, at the formal level, the following integral transform
\be\label{transform}
f(z)=\sum_{k\geq 0}f_k z^{-k-\frac 1 2}\mapsto \wh f (t):=\sum_{k\geq 0}\frac{f_k}{(2k)!}t^{2k}, \quad \ f(z)=\int\wh f(t)\e^{-t\sqrt{z}}\d t
\ee
for which
\be\label{diff}
2\wh{zf'(z)}= -\frac{\d}{\d t}(t\wh f(t)), \ \ \ \wh{z f(z)}=\frac{\d^2}{\d t^2}\wh f(t).
\ee
Hence, by \eqref{system} and \eqref{diff}, the formal series $\wh{\U_{11}}(t),\wh{\U_{12}}(t),\wh{\U_{21}}(t)$ satisfy the system
\be\label{systemtransformed}
\begin{cases}
-\frac{\d}{\d t}\left(t\wh{\U_{11}}(t)\right)=-2\frac{\d^2}{\d t^2}\wh{\U_{12}}(t)+2\wh{\U_{21}}(t)
\\
-\frac{\d}{\d t}\left(t\wh{\U_{12}}(t)\right)=-4\wh{\U_{11}}(t)-2\nu\wh{\U_{12}}(t)
\\
-\frac{\d}{\d t}\left(t\wh{\U_{21}}(t)\right)=4\frac{\d^2}{\d t^2}\wh{\U_{11}}(t)+2\nu\wh{\U_{21}}(t).
\end{cases}
\ee
Solving for $\wh{\U_{11}}(t)$ and $\wh{\U_{21}}(t)$ from the first two equations in \eqref{systemtransformed} we obtain
\begin{align}\label{relations}
\wh{\U_{11}}(t)&=\frac{1-2\nu}{4}\wh{\U_{12}}(t)+\frac t 4 \frac{\d}{\d t}\wh{\U_{12}}(t),
\\
\nonumber \wh{\U_{21}}(t)&=\frac{2\nu-1}{8}\wh{\U_{12}}(t)+\frac{2\nu-3}{8}t\frac{\d}{\d t}\wh{\U_{12}}(t)+\left(1-\frac{t^2}8\right)\frac{\d^2}{\d t^2}\wh{\U_{12}}(t)
\end{align}
and inserting this in the third equation in \eqref{systemtransformed} we obtain ODE
\be\label{laplace}
t \left(16-t^2\right)\frac{\d^3}{\d t^3}\wh{\U_{12}}(t)
+2 \left(16-3 t^2\right) \frac{\d^2}{\d t^2}\wh{\U_{12}}(t)
+\left(4 \nu^2-7\right) t \frac{\d}{\d t}\wh{\U_{12}}(t)
+ \left(4\nu^2-1\right) \wh{\U_{12}}(t)
=0.
\ee
Now, from the expansions \cite{AbSt1965}
\be
I_\a(x)\sim \frac{1}{\sqrt{2\pi x}}\e^{x}\sum_{k\geq 0}\frac{\left(\frac 1 2 -\nu\right)_k\left(\frac 1 2 +\nu\right)_k}{k!(2x)^k},\ \ \
K_\a(x)\sim \sqrt{\frac{\pi}{2x}}\e^{-x}\sum_{k\geq 0}\frac{(-1)^k\left(\frac 1 2 -\nu\right)_k\left(\frac 1 2 +\nu\right)_k}{k!(2x)^k}
\ee
we see that
\be
\U_{12}(z)=4I_{-\nu}(2\sqrt{z})K_{-\nu}(2\sqrt{z})=\frac{1}{\sqrt{z}}\left(1+\O\left(\frac{1}{z}\right)\right)
\ee
is a power series containing only negative odd powers of $\sqrt{z}$ and so, from \eqref{transform},
\be\label{beh}
\wh{\U_{12}}(t)=1+\O\left(t^2\right)
\ee
is a power series containing only positive even powers of $t$. Hence we are interested in even power series solutions $\wh{\U_{12}}(t)=1+\O\left(t^2\right)$ of the ODE \eqref{laplace}; by the Frobenius method it is possible to conclude that there exists exactly one such solution, which can be written in closed form in terms of the Gauss hypergeometric function
\be
\wh{\U_{12}}(t)={_2F_1}\left(\frac 1 2-\nu, \frac 1 2 +\nu; 1;\frac{t^2}{16}\right)=\sum_{k\geq 0}\frac{\left(\frac 1 2-\nu\right)_k\left(\frac 1 2+\nu\right)_k}{(k!)^2}\frac{t^{2k}}{16^k}.
\ee
Finally, recalling transformation \eqref{transform} we have
\be\label{UU}
\sqrt{z}\U_{12}(z)=\sum_{k\geq 0}\frac{\left(\frac 1 2-\nu\right)_k\left(\frac 1 2+\nu\right)_k(2k)!}{(k!)^2}\frac{z^{-k}}{16^k}
\ee
which simplifies to the (1,2)--entry in \eqref{U} by the identity $(2k)!=2^kk!(2k-1)!!$. The other entries of \eqref{U} are obtained by substituting \eqref{UU} into \eqref{relations}. \QED

Returning now to \eqref{s1}, we compute using \eqref{U}
\begin{align}
&\tr\left(\begin{bmatrix} -\frac{\nu}{2z} & \frac 1 z \\ 1 & \frac{\nu}{2z}
\end{bmatrix}\mathcal{U}(z;\nu)\right)=-\frac{\nu}{z}\U_{11}(z)+\frac{\U_{21}(z)}{z}+\U_{12}(z)\nonumber \\
&\qquad =-\frac{\nu}{2}\sum_{k\geq 0}\frac{(2k-1)!!}{8^kk!}\left(\frac 1 2-\nu\right)_{k+1}\left(\frac 1 2+\nu\right)_kz^{-1-k}\nonumber\\
&\qquad \phantom{{}=}+\sum_{k\geq 0}\frac{(2k-1)!!}{8^{k}k!}\left[-\left(\frac 1 2 -\nu\right)_{k+1}\left(\frac 1 2+\nu\right)_{k-1}+\left(\frac 1 2-\nu\right)_k\left(\frac 1 2+\nu\right)_k\right]z^{-k} \nonumber \\
&\qquad=-\frac{\nu}{2}\sum_{k\geq 1}\frac{(2k-3)!!}{8^{k-1}(k-1)!}\left(\frac 1 2-\nu\right)_{k}\left(\frac 1 2+\nu\right)_{k-1}z^{-k}+2\nonumber \\
&\qquad \phantom{{}=}
-\sum_{k\geq 1}\frac{(2k-1)!!}{8^kk!}\left(\frac 1 2-\nu\right)_{k-1}\left(\frac 1 2+\nu\right)_kz^{-k}
\nonumber \\
&\qquad=2-\sum_{k\geq 1}\frac{(2k-3)!!}{2^{3k-1}k!}\left(\frac 1 2-\nu\right)_k\left(\frac 1 2+\nu\right)_kz^{-k}
\end{align}
hence \eqref{s1} gives
\be
S_1(z;\nu)=\sum_{\ell\geq 0}z^{-1-\ell}\left.\frac{\pa\tau(\t;\nu)}{\pa t_\ell}\right|_{\t=0}=\sum_{k\geq 1}\frac{(2k-3)!!}{2^{3k-1}k!}\left(\frac 1 2-\nu\right)_k\left(\frac 1 2+\nu\right)_kz^{-k}
\ee
from which \eqref{onepoint} follows by the change of variable $k=1+\ell$.

\subsubsection{n--point function}

We first consider the two--point function; apply $\sum\limits_{\ell_2\geq 0}z^{-1-\ell_2}_2\frac{\pa}{\pa t_{\ell_2}}$ on \eqref{s1t} to get
\begin{align}
&\sum_{\ell_2\geq 0}z_1^{-\ell_1-1}z_2^{-\ell_2-1}\frac{\pa^2\log\tau(\t)}{\pa t_{\ell_1}\pa t_{\ell_2}}
\\
&\qquad =
-\sum_{\ell_2\geq 0}z_2^{-\ell_2-1}\tr\left(\sqrt{z_1}\Omega_{\ell_2}'(z_1;\t)\Psi(z_1;\t)\s_3\Psi^{-1}(z_1;\t)\right)
-2\sum_{\ell_2\geq 0}z^{-1-\ell_2}_2\sqrt{z_1}\frac{\pa}{\pa t_{\ell_2}}\vartheta'(z_1;\t).
\nonumber 
\end{align}
The second term is easily computed as
\be
-2\sum_{\ell_2\geq 0}z^{-1-\ell_2}_2\sqrt{z_1}\frac{\pa}{\pa t_{\ell_2}}\vartheta'(z_1;\t)=-2\sum_{\ell_2\geq 0}\frac{2\ell_2+1}{2}z_1^{\ell_2}z_2^{-1-\ell_2}=-\frac{z_1+z_2}{(z_1-z_2)^2}.
\ee
For the first one we introduce
\be
\label{defR}
\mathcal{R}(z;\t):=\sqrt{z}\Psi(z;\t)\s_3\Psi^{-1}(z;\t)
\ee
and we rewrite from \eqref{Omega}
\be\label{Omegaaaa}
\Omega_\ell(z;\t)=\res{w=\infty}\frac{\mathcal{R}(w;\t)w^\ell}{w-z}\d w.
\ee
Consequently we obtain
\begin{align}
&\sum_{\ell_2\geq 0}z_2^{-\ell_2-1}\tr\left(\sqrt{z_1}\Omega_{\ell_2}'(z_1;\t)\Psi(z_1;\t)\s_3\Psi^{-1}(z_1;\t)\right)\nonumber
\\ 
&\qquad=-\sum_{\ell_2\geq 0}z_2^{-\ell_2-1}\tr\left(\res{w=\infty}\frac{\mathcal{R}(w;\t)w^{\ell_2}}{(w-z_1)^2}\mathcal{R}(z_1;\t)\right)\d w
=\tr\left(\frac{\mathcal{R}(z_2;\t)\mathcal{R}(z_1;\t)}{(z_2-z_1)^2}\right)
\end{align}
and, using  Lemma \eqref{lemmaU}, evaluation at $\t=0$ gives
\be\label{evaluationattzero}
\left.\mathcal{R}(z;\t)\right|_{\t=0}=\sqrt{z}\,\Xi(z)\s_3\Xi^{-1}(z)=\mathcal{U}(z;\nu)
\ee
and \eqref{npoint} is proven for $n=2$.

To prove \eqref{npoint} for arbitrary $n\geq 3$ we state the following Lemma.

\begin{lemma}\label{lemmanpoint}
For all $n\geq 2$ we have
\begin{align}
\nonumber
&\sum_{\ell_1,...,\ell_n\geq 0}z_1^{-1-\ell_1}\cdots z_n^{-1-\ell_n}\frac{\pa^n\log\tau(\t)}{\pa t_{\ell_1}\cdots\pa t_{\ell_n}}
\\
\label{npointt}
&\qquad =\frac{(-1)^{n-1}}{n}\sum_{\iota\in \mathfrak{S}_n}\frac{\tr\left(\mathcal{R}(z_1;\t)\cdots\mathcal{R}(z_n;\t)\right)}{(z_{\iota_1}-z_{\iota_2})\cdots(z_{\iota_{n-1}}-z_{\iota_n})(z_{\iota_n}-z_{\iota_1})}
-\frac{z_1+z_2}{(z_1-z_2)^2}\delta_{n,2}.
\end{align}
\end{lemma}
\noindent {\bf Proof.} The proof is given by induction on $n\geq 2$; the induction base $n=2$ has been proven above. Assume \eqref{npointt} holds true for some $n\geq 2$ then, writing $\mathcal{R}(z):=\mathcal{R}(z;\t)$ for short,
\begin{align}
&\sum_{\ell_1,...,\ell_{n+1}\geq 0}z_1^{-1-\ell_1}\cdots z_{n+1}^{-1-\ell_{n-1}}\frac{\pa^{n+1}\log\tau(\t)}{\pa t_{\ell_1}\cdots \pa t_{\ell_{n+1}}}\nonumber
\\
&\qquad=\frac{(-1)^{n-1}}{n}\sum_{\ell_{n+1}\geq 0}\sum_{\iota\in \mathfrak{S}_n}z_{n+1}\frac{\pa}{\pa t_{\ell_{n+1}}}\frac{\tr\left(\mathcal{R}(z_1)\cdots\mathcal{R}(z_n)\right)}{(z_{\iota_1}-z_{\iota_2})\cdots(z_{\iota_{n-1}}-z_{\iota_n})(z_{\iota_n}-z_{\iota_1})}.
\label{lasst}
\end{align}
Using $\frac{\pa}{\pa t_\ell}\mathcal{R}(z)=[\Omega_\ell(z),\mathcal{R}(z)]$ it can be derived from \eqref{Omegaaaa} that
\be
\sum_{\ell\geq 0}z^{-1-\ell}\frac{\pa}{\pa t_\ell}\mathcal{R}(z')=\frac{[\mathcal{R}(z),\mathcal{R}(z')]}{z'-z}
\ee
and so we rewrite \eqref{lasst} as
\begin{align}
&\frac{(-1)^{n-1}}{n}\sum_{\iota\in \mathfrak{S}_n}\sum_{j=1}^n\frac{\tr\left(\mathcal{R}(z_1)\cdots\left(\mathcal{R}(z_{n+1})\mathcal{R}(z_{\iota_{j}})-\mathcal{R}(z_{\iota_{j}})\mathcal{R}(z_{n+1})\right)\cdots\mathcal{R}(z_n)\right)}{(z_{\iota_1}-z_{\iota_2})\cdots(z_{\iota_{n-1}}-z_{\iota_n})(z_{\iota_n}-z_{\iota_1})(z_{\iota_j}-z_{n+1})}\nonumber
\\
&\qquad=\frac{(-1)^{n-1}}{n}\sum_{\iota\in \mathfrak{S}_n}\sum_{j=1}^n\frac{\tr\left(\mathcal{R}(z_{\iota_1})\cdots\mathcal{R}(z_{\iota_{j-1}})\mathcal{R}(z_{n+1})\mathcal{R}(z_{\iota_j})\cdots\mathcal{R}(z_{\iota_n})\right)}{(z_{\iota_1}-z_{\iota_2})\cdots(z_{\iota_{n-1}}-z_{\iota_n})(z_{\iota_n}-z_{\iota_1})}\nonumber
\\ &\qquad\phantom{{}=}\times\left(\frac{1}{z_{\iota_j}-z_{n+1}}-\frac{1}{z_{\iota_{j+1}}-z_{n+1}}\right)\nonumber
\\
&\qquad=\frac{(-1)^n}{n}\sum_{\iota\in \mathfrak{S}_n}\sum_{j=1}^n\frac{\tr\left(\mathcal{R}(z_{\iota_1})\cdots\mathcal{R}(z_{\iota_{j-1}})\mathcal{R}(z_{n+1})\mathcal{R}(z_{\iota_j})\cdots\mathcal{R}(z_{\iota_n})\right)}{(z_{\iota_1}-z_{\iota_2})\cdots(z_{\iota_j}-z_{n+1})(z_{n+1}-z_{\iota_{j+1}})\cdots(z_{\iota_{n-1}}-z_{\iota_n})(z_{\iota_n}-z_{\iota_1})}
\nonumber
\\
&\qquad=\frac{(-1)^n}{n+1}\sum_{\iota\in \mathfrak{S}_{n+1}}\frac{\tr\left(\mathcal{R}(z_{\iota_1})\cdots\mathcal{R}(z_{\iota_{n+1}})\right)}{(z_{\iota_1}-z_{\iota_2})\cdots(z_{\iota_{n}}-z_{\iota_{n+1}})(z_{\iota_{n+1}}-z_{\iota_1})}
\end{align}
where we have used the cyclic property of the trace.\QED

Finally \eqref{npoint} follows by evaluating \eqref{npointt} at $\t=0$ using \eqref{evaluationattzero}, and the proof of Thm. \ref{thmcor} is complete.

\subsection{Proof of Thm. \ref{thmvir}}\label{secproofvir}

Here we prove Thm. \ref{thmvir}; hereafter we drop the explicit notation of dependence on $z,\t,\nu$ and denote
\be
\wt t_\ell:=t_\ell-2\delta_{\ell,0}, \ \ \ \pa_\ell:=\frac{\pa}{\pa\wt t_\ell}=\frac{\pa}{\pa t_\ell}.
\ee

\subsubsection{Preliminaries}

We collect here some simple results that will be needed below.

\begin{lemma}
The following identity holds true for all $k\geq 0$;
\be\label{mainid}
\res{z=\infty}\tr\left(zA'\Psi\s_3\Psi^{-1}\sqrt{z}^{2k+1}\right)\d z+\frac{2k+3}{2}\pa_k\log\tau=0.
\ee
\end{lemma}
\noindent {\bf Proof.} 
The (formal or not) residue of a total differential vanishes, hence
\be 
\res{z=\infty}\tr\left(\Psi'\s_3\Psi^{-1}\sqrt{z}^{2k+3}\right)'\d z=0
\ee
and computing the left hand side using $\Psi'=A\Psi$ we have
\begin{align}
&\res{z=\infty}\tr\left((A\Psi)'\s_3\Psi^{-1}\sqrt{z}^{2k+3}-A\Psi\s_3\Psi^{-1}\Psi'\Psi^{-1}\sqrt{z}^{2k+3}+\frac{2k+3}{2}\Psi'\s_3\Psi^{-1}\sqrt{z}^{2k+1}\right)\d z\nonumber
\\
&\qquad =\res{z=\infty}\tr\left(A'\Psi\s_3\Psi^{-1}+\cancel{A^2\Psi\s_3\Psi^{-1}}-\cancel{A\Psi\s_3\Psi^{-1}A}\right)\sqrt{z}^{2k+3}\d z+\frac{2k+3}{2}\pa_k\log\tau
\end{align}
where the two terms indicated cancel out thanks to the cyclic property of the trace.
\QED

\begin{lemma}\label{lemmader}
The following formul\ae\  hold true, for all $a,b,c\geq 0$;
\begin{align}\label{secondder}
\pa_b\pa_c\log\tau&=\res{z=\infty}\tr\left(\Omega_b'\Psi\s_3\Psi^{-1}\sqrt{z}^{2c+1}\right)\d z,
\\
\label{thirdder}
\pa_a\pa_b\pa_c\log\tau&=\res{z=\infty}\tr\left(\left(\pa_a\Omega_b'+[\Omega_b',\Omega_a]\right)\Psi\s_3\Psi^{-1}\sqrt{z}^{2c+1}\right)\d z.
\end{align}
\end{lemma}
\noindent {\bf Proof.} We start from the definition \eqref{isomonodromicinfty}
\be
\pa_c\log\tau=\res{z=\infty}\tr\left(\Psi'\s_3\Psi^{-1}\sqrt{z}^{2c+1}\right)\d z
\ee
and applying $\pa_b$ using $\pa_b\Psi=\Omega_b\Psi$ we get
\begin{align}
\pa_b\pa_c\log\tau&=\res{z=\infty}\tr\left(\left(\Omega_b\Psi\right)'\s_3\Psi^{-1}\sqrt{z}^{2c+1}-\Psi'\s_3\Psi^{-1}\Omega_b\sqrt{z}^{2c+1}\right)\d z\nonumber
\\
&=\res{z=\infty}\tr\left(\Omega_b'\Psi\s_3\Psi^{-1}\sqrt{z}^{2c+1}+\cancel{\Omega_b\Psi'\s_3\Psi^{-1}\sqrt{z}^{2c+1}}-\cancel{\Psi'\s_3\Psi^{-1}\Omega_b\sqrt{z}^{2c+1}}\right)\d z
\end{align}
where the two terms cancel due to the cyclic property of the trace; \eqref{secondder} is proven. Now apply $\pa_a$ to \eqref{secondder} to obtain
\be
\pa_a\pa_b\pa_c\log\tau=\res{z=\infty}\tr\left(\left(\left(\pa_a\Omega_b'\right)\Psi\s_3\Psi^{-1}+\Omega_b'\Omega_a\Psi\s_3\Psi^{-1}-\Omega_b'\Psi\s_3\Psi^{-1}\Omega_a\right)\sqrt{z}^{2c+1}\right)\d z
\ee
which simplifies to \eqref{thirdder}, once again thanks to the cyclic property of the trace.
\QED

As a last preliminary,  let us use the expansion
\be\label{expansionY}
Y(z;\t)=\1+\begin{bmatrix}
a & a \\ -a & -a
\end{bmatrix}z^{-\frac{1}{ 2}}+\begin{bmatrix}
b & c \\ c & b
\end{bmatrix}z^{-1}+\begin{bmatrix}d & e \\ -e & -d
\end{bmatrix}z^{-\frac 3 2}+\begin{bmatrix}f & g \\ g & f\end{bmatrix}z^{-2}+\O\left(z^{-\frac 5 2}\right)
\ee
with $a=a(\t)$,..., $g=g(\t)$, to compute
\begin{align}
\label{Omega0}
\Omega_0&=
\begin{bmatrix}
-2a & -1 \\ -z-2c & 2a
\end{bmatrix},
\\
\label{Omega1}
\Omega_1&=
\begin{bmatrix}
 2 (a b-ac-e)-2 a z & 4 a^2+2 c-z \\
 2 (a e-ad- c^2+ b c- g)-2 z c-z^2 & 2 (-a b+a c+e)+2 a z 
\end{bmatrix},
\end{align}
and, by direct use of \eqref{isomonodromicinfty} we also find
\begin{align}
\label{pa0}
\pa_0\log\tau&=2a,
\\
\label{pa1}
\pa_1\log\tau&=-4ab+3d+e.
\end{align}

\subsubsection{Proof of $L_0\tau=0$}

We compute from \eqref{A}
\be\label{zA'}
zA'=z\left(\frac{\s_3}{4z^2}+\frac{1}{z}\sum_{\ell\geq 0}\frac{2\ell+1}{2}\wt t_\ell\Omega'_\ell-\frac{1}{z^2}\sum_{\ell\geq 0}\frac{2\ell+1}{2}\wt t_\ell\Omega_\ell\right)=\sum_{\ell\geq 0}\frac{2\ell+1}{2}\wt t_\ell\Omega'_\ell-A.
\ee
Substitution in \eqref{mainid} shows that for all $k\geq 0$ we have
\begin{align}
0&=\sum_{\ell\geq 0}\frac{2\ell+1}{2}\wt t_\ell\res{z=\infty}\tr(\Omega'_\ell\Psi\s_3\Psi\sqrt{z}^{2k+1})\d z-\res{z=\infty}\tr\left(A\Psi\s_3\Psi^{-1}\sqrt{z}^{2k+1}\right)+\frac{2k+3}{2}\pa_k\log\tau\nonumber
\\
&=\sum_{\ell\geq 0}\frac{2\ell+1}{2}\wt t_\ell\pa_\ell\pa_k\log\tau+\frac{2k+1}{2}\pa_k\log\tau=\pa_k\left( \frac{L_0\tau}{\tau}\right)
\end{align}
where we use \eqref{secondder} and the fact that $A\Psi=\Psi'$; the last identity implies $\frac{L_0\tau}{\tau}=C$ for some constant $C$; evaluation at $t_\ell=0$, i.e. $\wt t_\ell=-2\delta_{\ell,0}$, using the definition of $L_0$ in \eqref{vir} shows that
\be
C=\left. \frac{L_0\tau}{\tau}\right|_{\t=0}=\left.-\pa_0\log\tau\right|_{\t=0}+\frac{1-4\nu^2}{16}=-\frac{1-4\nu^2}{16}+\frac{1-4\nu^2}{16}=0
\ee
where we use $\left.\pa_0\log\tau\right|_{\t=0}=\frac{1-4\nu^2}{16}$, which follows either by the explicit formula \eqref{onepoint} or by \eqref{taux} with $x=0$. Therefore $L_0\tau=0$.

\begin{remark}
The constraint $L_0\tau=0$ follows also from the dilation covariance of the RHP \ref{probleminfty}. Concretely, the matrix $\Psi(\e^{u} z;\t)$ ($u\in\R$) satisfies the same jump condition as $\Psi(z;\t)$, as the latter has been defined in \eqref{Psiinfty} and satisfies a jump condition with matrices independent of $z,\t$; further we have the boundary behaviour
\be
\Psi(\e^uz)\sim \e^{-\frac{u}{4}\s_3}z^{-\frac{\s_3}{4}}G\left(\1+\begin{bmatrix}a(\t) & a(\t) \\ -a(\t) &-a(\t)\end{bmatrix}\e^{-\frac{u}{2}}z^{-\frac 1 2}+\O\left(z^{-1}\right)\right)\e^{-\vartheta(z;\t(u))\s_3}, \ \ \ z\to\infty
\ee
where $t_\ell(u):=\e^{\frac{2\ell+1}{2}u}t_\ell$. It follows that $\e^{\frac{u}{4}\s_3}\G(\e^uz;\t(-u))$ solves RHP \ref{probleminfty}, the solution of which is unique, hence
\be
\G(z;\t)=\e^{\frac{u}{4}\s_3}\G(\e^uz;\t(-u)).
\ee
Therefore, for all $k\geq 0$ we have
\be
\res{z=\infty}\tr\left(\G^{-1}(z;\t)\G'(z;\t)\s_3\sqrt{z}^{2k+1}\right)=\res{z=\infty}\tr\left(\G^{-1}(\e^uz;\t(-u))\G'(\e^uz;\t(-u))\s_3\sqrt{z}^{2k+1}\right)
\ee
and the last expression does not depend on $u$ by construction; setting the first variation in $u$ equal to zero we recover $\pa_k\left(\frac{L_0\tau}{\tau}\right)=0$ for all $k\geq 0$, from which we can derive $L_0\tau=0$ as above.

Note that due to the special point $z=0$, RHP \ref{probleminfty} does not have a translation covariance property.
\end{remark}

\subsubsection{Proof of $L_1\tau=0$}
As a consequence of the recursion
\be\label{recursion}
z\Omega_\ell=\Omega_{\ell+1}-\left(\Omega_{\ell+1}\right)_0\Rightarrow z\Omega_\ell'=\Omega_{\ell+1}'-\Omega_\ell
\ee
where $()_0$ denotes the constant term in $z$, we multiply \eqref{zA'} by $z$ to get
\begin{align}\label{zzA'}
z^2A'&=\sum_{\ell\geq 0}\frac{2\ell+1}{2}\wt t_\ell z\Omega_\ell'-zA=\sum_{\ell\geq 0}\frac{2\ell+1}{2}\wt t_\ell \Omega_{\ell+1}'-\sum_{\ell\geq 0}\frac{2\ell+1}{2}\wt t_\ell \Omega_{\ell}-zA\nonumber
\\
&=\sum_{\ell\geq 0}\frac{2\ell+1}{2}\wt t_\ell \Omega_{\ell'+1}-\underbrace{\left(\sum_{\ell\geq 0}\frac{2\ell+1}{2}\wt t_\ell \Omega_{\ell}-\frac{\s_3}{4}\right)}_{=zA}-\frac{\s_3}{4}-zA=\sum_{\ell\geq 0}\frac{2\ell+1}{2}\wt t_\ell \Omega_{\ell+1}'-2zA-\frac{\s_3}{4}
\end{align}
and we use \eqref{mainid} with $k\mapsto k+1$:
\begin{align}\label{eql1}
0&=\res{z=\infty}\tr\left(z^2A'\Psi\s_3\Psi^{-1}\sqrt{z}^{2k+1}\right)\d z+\frac{2k+5}{2}\pa_k\log\tau\nonumber
\\
&=\sum_{\ell\geq 0}\frac{2\ell+1}{2}\wt t_\ell\res{z=\infty}\tr(\Omega'_{\ell+1}\Psi\s_3\Psi\sqrt{z}^{2k+1})\d z-\res{z=\infty}\tr\left(2zA\Psi\s_3\Psi^{-1}\sqrt{z}^{2k+1}\right)\nonumber
\\
&\phantom{{}=}+\frac{2k+5}{2}\pa_{k+1}\log\tau-\frac{1}{4}\res{z=\infty}\tr\left(\s_3\Psi\s_3\Psi^{-1}\sqrt{z}^{2k+1}\right)\d z\nonumber
\\
&=\sum_{\ell\geq 0}\frac{2\ell+1}{2}\wt t_\ell\pa_{\ell+1}\pa_k\log\tau+\frac{2k+1}{2}\pa_{k+1}\log\tau-\frac{1}{4}\res{z=\infty}\tr\left(\s_3\Psi\s_3\Psi^{-1}\sqrt{z}^{2k+1}\right)\d z
\end{align}
where we have used \eqref{secondder} and $A\Psi=\Psi'$.
\begin{lemma}\label{lemmal1}
We have
\be\label{ehmboh}
-\res{z=\infty}\tr\left(\s_3\Psi\s_3\Psi^{-1}\sqrt{z}^{2k+1}\right)\d z=\pa_k\left(\frac{\pa_0^2\tau}{\tau}\right).
\ee
\end{lemma}

\noindent {\bf Proof of Lemma \ref{lemmal1}.} Note that
\begin{align}
\pa_k\left(\frac{\pa_0^2\tau}{\tau}\right)&=\pa_k\left(\pa_0^2\log\tau+(\pa_0\log\tau)^2\right)\nonumber
\\
&=\pa_k\pa_0^2\log\tau+2(\pa_0\log\tau)(\pa_k\pa_0\log\tau)=\pa_k\pa_0^2\log\tau+4a \pa_k\pa_0\log\tau
\end{align}
where we have used \eqref{pa0} in the last step. Using Lemma \ref{lemmader} we obtain
\be
\pa_k\pa_0^2\log\tau+4a \pa_k\pa_0\log\tau=\res{z=\infty}\tr\left(\left(\pa_0\Omega_0'+[\Omega_0',\Omega_0]+4a \Omega_0'\right)\Psi\s_3\Psi^{-1}\sqrt{z}^{2k+1}\right)\d z
\ee
and the statement \eqref{ehmboh} boils down to the identity
\be
\cancel{\pa_0\Omega_0'}+[\Omega_0',\Omega_0]+4a \Omega_0'=-\s_3
\ee
which is easily checked using \eqref{Omega0}. \QED

Back to the proof of $L_1\tau=0$, we see from the last line of \eqref{eql1} together with Lemma \ref{lemmal1} that we have proven $\pa_k\left(\frac{L_1\tau}{\tau}\right)=0$ for all $k\geq 0$. Hence $L_1\tau=C\tau$ for some constant $C$; evaluation at $\t=(0,0,...)$ shows that $C=0$, e.g. by using \eqref{tau}, and so $L_1\tau=0$.

\subsubsection{Proof of $L_2\tau=0$}
Using the recursion \eqref{recursion} we see that
\be
z\Omega'_{\ell+1}=\Omega_{\ell+2}'-\Omega_{\ell+1}=\Omega_{\ell+2}'-z\Omega_\ell-\left(\Omega_{\ell+1}\right)_0
\ee
where again we denote $()_0$ the constant term in $z$; we then compute from \eqref{zzA'}
\begin{align}\label{zzzA'}
z^3A'&=\sum_{\ell\geq 0}\frac{2\ell+1}{2}\wt t_\ell z\Omega'_{\ell+1}-2z^2A-\frac{\s_3}{4}z\nonumber
\\
&=\sum_{\ell\geq 0}\frac{2\ell+1}{2}\wt t_\ell\Omega'_{\ell+2}-z\sum_{\ell\geq 0}\frac{2\ell+1}{2}\wt t_\ell\Omega_{\ell}-\sum_{\ell\geq 0}\frac{2\ell+1}{2}\wt t_\ell\left(\Omega_{\ell+1}\right)_0-2z^2A-\frac{\s_3}{4}z\nonumber
\\
&=\sum_{\ell\geq 0}\frac{2\ell+1}{2}\wt t_\ell\Omega'_{\ell+2}-\sum_{\ell\geq 0}\frac{2\ell+1}{2}\wt t_\ell\left(\Omega_{\ell+1}\right)_0-3z^2A-\frac{\s_3}{2}z.
\end{align}

\begin{lemma}\label{lemmal2}
We have the identity
\be
-\sum_{\ell\geq 0}\frac{2\ell+1}{2}\wt t_\ell\left(\Omega_{\ell+1}\right)_0=\begin{bmatrix}
-b+c & -a \\ \frac{3}{2}(d-e) &-b-c
\end{bmatrix}
\ee
where $a=a(\t),...,e=e(\t)$ are as in \eqref{expansionY}.
\end{lemma}

\noindent {\bf Proof of Lemma \ref{lemmal2}.} Since  $(z^2\Psi)'$ satisfies the same jump condition  as $\Psi$ along $\S$, it follows that the ratio $(z^2\Psi)'\Psi^{-1}$ is an entire matrix--valued function; indeed from \eqref{growthinftyzero} we see that this ratio is analytic also at $z=0$. Since this ratio has polynomial growth at $z=\infty$, see \eqref{growthinftyinfty}, we conclude that $(z^2\Psi)'\Psi^{-1}$ is actually a polynomial, which coincides with the polynomial part of its expansion at $z=\infty$;
\begin{align}\label{booooooooh}
(z^2\Psi)'\Psi^{-1}&=\left(2z\1-z^2\frac{\s_3}{4z}+z^2z^{-\frac{\s_3}{4}}GY'Y^{-1}G^{-1}z^{\frac{\s_3}{4}}+z^2\Psi\vartheta'\s_3\Psi^{-1}\right)_+\nonumber
\\
&=2z\1-z\frac{\s_3}{4}+\left(z^2z^{-\frac{\s_3}{4}}GY'Y^{-1}G^{-1}z^{\frac{\s_3}{4}}\right)_++\sum_{\ell\geq 0}\frac{2\ell+1}{2}\wt t_\ell\Omega_{\ell+1}.
\end{align}
However, it is trivial to compute $(z^2\Psi)'\Psi^{-1}=2z\1+z^2 A$, which has no constant term in $z$. Therefore also the constant term in $z$ in \eqref{booooooooh} vanishes and hence
\be 
-\left(\sum_{\ell\geq 0}\frac{2\ell+1}{2}\wt t_\ell\Omega_{\ell+1}\right)_0=\left(z^2z^{-\frac{\s_3}{4}}GY'Y^{-1}G^{-1}z^{\frac{\s_3}{4}}\right)_0=\begin{bmatrix}
-b+c & -a \\ \frac{3}{2}(d-e) &-b-c
\end{bmatrix}
\ee
and the Lemma is proven. \QED

Back to the proof of $L_2\tau=0$, we obtain from \eqref{zzzA'} together with Lemma \ref{lemmal2}
\be
\renewcommand*{\arraystretch}{1.25}
z^3A'=\sum_{\ell\geq 0}\frac{2\ell+1}{2}\wt t_\ell\Omega'_{\ell+2}-3z^2A+\begin{bmatrix}
-\frac{z}{2}-b+c & -a \\ \frac{3}{2}(d-e) &\frac{z}{2}-b-c
\end{bmatrix}
\ee
and inserting this expression in \eqref{mainid} with $k\mapsto k+2$ we have
\begin{align}\label{meh}
0&=\res{z=\infty}\tr\left(z^3A'\Psi\s_3\Psi^{-1}\sqrt{z}^{2k+1}\right)\d z+\frac{2k+7}{2}\pa_{k+2}\log\tau\nonumber
\\
&=\sum_{\ell\geq 0}\frac{2\ell+1}{2}\wt t_\ell\res{z=\infty}\tr\left(\Omega'_{\ell+2}\Psi\s_3\Psi^{-1}\sqrt{z}^{2k+1}\right)\d z
-3\res{z=\infty}\left(z^3A\Psi\s_3\Psi^{-1}\sqrt{z}^{2k+1}\right)\d z\nonumber
\\
&\phantom{{}=}
\renewcommand*{\arraystretch}{1.25}
+\res{z=\infty}\tr\left(\begin{bmatrix}
-\frac{z}{2}-b+c & -a \\ \frac{3}{2}(d-e) &\frac{z}{2}-b-c\end{bmatrix}\Psi\s_3\Psi^{-1} \sqrt{z}^{2k+1}\right)\d z+\frac{2k+7}{2}\pa_{k+2}\log\tau\nonumber
\\
&=\sum_{\ell\geq 0}\frac{2\ell+1}{2}\wt t_\ell\pa_k\pa_{\ell+2}\log\tau+\frac{2k+1}{2}\pa_{k+2}\log\tau\nonumber
\\
&\phantom{{}=}
\renewcommand*{\arraystretch}{1.25}
+\res{z=\infty}\tr\left(\begin{bmatrix}
-\frac{z}{2}-b+c & -a \\ \frac{3}{2}(d-e) &\frac{z}{2}-b-c\end{bmatrix}\Psi\s_3\Psi^{-1} \sqrt{z}^{2k+1}\right)\d z.
\end{align}
The final part is the computation of the last term in the above equation. This is done in the following Lemma.

\begin{lemma}\label{lemmal22}
We have
\be
\renewcommand*{\arraystretch}{1.25}
\res{z=\infty}\tr\left(\begin{bmatrix}
-\frac{z}{2}-b+c & -a \\ \frac{3}{2}(d-e) &\frac{z}{2}-b-c\end{bmatrix}\Psi\s_3\Psi^{-1} \sqrt{z}^{2k+1}\right)\d z
=
\pa_k\left(\frac{\pa_0\pa_1\tau}{2\tau}\right).
\ee
\end{lemma}

\noindent {\bf Proof of Lemma \ref{lemmal22}.} Note that
\begin{align}
\pa_k\left(\frac{\pa_0\pa_1\tau}{\tau}\right)
&=\pa_k\left(\pa_0\pa_1\log\tau+\left(\pa_0\log\tau\right)\left(\pa_1\log\tau\right)\right)\nonumber
\\
&=\pa_k\pa_0\pa_1\log\tau+\left(\pa_k\pa_0\log\tau\right)\left(\pa_1\log\tau\right)+\left(\pa_0\log\tau\right)\left(\pa_k\pa_1\log\tau\right)\nonumber
\\
&=\pa_k\pa_0\pa_1\log\tau+(-4ab+3d+e)\left(\pa_k\pa_0\log\tau\right)+2a\left(\pa_k\pa_1\log\tau\right)
\end{align}
where we have used \eqref{pa0} and \eqref{pa1}. Using Lemma \ref{lemmader} and the explicit expressions \eqref{Omega0} and \eqref{Omega1} we obtain
\begin{align}
\pa_k\left(\frac{\pa_0\pa_1\tau}{2\tau}\right)&=\frac{1}{2}\res{z=\infty}\tr\left(\left(\pa_1\Omega_0'+[\Omega_0',\Omega_1]+(-4ab+3d+e)\Omega_0'+2a\Omega_1'\right)\Psi\s_3\Psi^{-1}\sqrt{z}^{2k+1}\right)\d z\nonumber
\\
&=\res{z=\infty}\tr\left(\begin{bmatrix}-\frac z 2 + c & -a \\ \frac{3}{2}(-d+e) & \frac z 2 -c
\end{bmatrix}
\Psi\s_3\Psi^{-1}\sqrt{z}^{2k+1}\right)\d z
\end{align}
and the proof is complete, as $\tr\left(\begin{bmatrix}
-b & 0 \\ 0 &-b\end{bmatrix}\Psi\s_3\Psi^{-1}\right)=-b\ \tr\left(\Psi\s_3\Psi^{-1}\right)=-b\ \tr(\s_3)=0$.
\QED

From the last line of \eqref{meh} combined with Lemma \ref{lemmal22} we obtain $\pa_k\left(\frac{L_2\tau}{\tau}\right)=0$, for all $k\geq 0$. It follows that $L_2\tau=C\tau$ for some integration constant $C$; evaluation at $\t=(0,0,...)$ shows that $C=0$, e.g. by using \eqref{tau}, and so $L_2\tau=0$.

\subsubsection{Proof of Thm. \ref{thmvir}}

We have proven $L_n\tau=0$ for $n=0,1,2$. It remains to show that $L_{n+1}\tau=0$ for $n\geq 2$. The proof is given by induction on $n\geq 2$: assume that $L_n\tau=0$ for some $n\geq 2$, then exploiting the Virasoro commutation relation \eqref{virasorocomm} we have
\be
L_{n+1}\tau=\frac{1}{n-1}\left(L_nL_1\tau-L_1L_n\tau\right)=0
\ee
and the proof of Thm. \ref{thmvir} is complete.

\appendix

\section{Tables of low genus $n$-point intersection numbers ($n=2,3,4$)}\label{table}

We introduce the notation 
\be
\langle\Theta,\tau_{\ell_1}\cdots\tau_{\ell_n}\rangle:=\int_{\overline{\mathcal{M}_{g,n}}}\Theta_{g,n} \psi^{\ell_1}_1 \cdots \psi_n^{\ell_n}
\ee
where $n\geq 1$ and the genus $g$ is found from the dimensional constraint as
\be
g=\ell_1+\cdots+\ell_n+1.
\ee
Below we list some intersection numbers $\langle\Theta,\tau_{\ell_1}\cdots\tau_{\ell_n}\rangle$ for $n=2,3,4$ and $1\leq \ell_1\leq\cdots \leq \ell_n$; insertions of arbitrary positive powers $\tau_0$ are not considered, as the corresponding intersection numbers can be computed by the relations
\be
\langle\Theta,\tau_0^k\tau_{\ell_1}\cdots\tau_{\ell_n}\rangle=(2g-2+n)_k\langle\Theta,\tau_{\ell_1}\cdots\tau_{\ell_n}\rangle, \qquad \langle\Theta,\tau_0\rangle=\frac 1 8
\ee
which follow from the Virasoro constraint $L_0\tau=0$.

\paragraph*{Two-point intersection numbers, $1\leq \ell_1\leq \ell_2\leq 10$}
\begin{tiny}
\begin{align*}
&\left\langle \Theta ,\tau _1^2\right\rangle =\frac{63}{512} && \left\langle \Theta ,\tau _1 \tau _2\right\rangle =\frac{8625}{32768} \\ & \left\langle \Theta ,\tau _2^2\right\rangle =\frac{125565}{131072} && \left\langle \Theta ,\tau _1 \tau _3\right\rangle =\frac{44835}{65536} \\ & \left\langle \Theta ,\tau _2 \tau _3\right\rangle =\frac{7949025}{2097152} && \left\langle \Theta ,\tau _3^2\right\rangle =\frac{178066035}{8388608} \\ & \left\langle \Theta ,\tau _1 \tau _4\right\rangle =\frac{8831025}{4194304} && \left\langle \Theta ,\tau _2 \tau _4\right\rangle =\frac{553978845}{33554432} \\ & \left\langle \Theta ,\tau _3 \tau _4\right\rangle =\frac{266956944345}{2147483648} && \left\langle \Theta ,\tau _4^2\right\rangle =\frac{8093029715505}{8589934592} \\ & \left\langle \Theta ,\tau _1 \tau _5\right\rangle =\frac{125893845}{16777216} && \left\langle \Theta ,\tau _2 \tau _5\right\rangle =\frac{169880880015}{2147483648} \\ & \left\langle \Theta ,\tau _3 \tau _5\right\rangle =\frac{1655391889305}{2147483648} && \left\langle \Theta ,\tau _4 \tau _5\right\rangle =\frac{1009001583045225}{137438953472} \\ & \left\langle \Theta ,\tau _5^2\right\rangle =\frac{38605283045457975}{549755813888} && \left\langle \Theta ,\tau _1 \tau _6\right\rangle =\frac{65335475205}{2147483648} \\ & \left\langle \Theta ,\tau _2 \tau _6\right\rangle =\frac{1782725109165}{4294967296} && \left\langle \Theta ,\tau _3 \tau _6\right\rangle =\frac{349269710865075}{68719476736} \\ & \left\langle \Theta ,\tau _4 \tau _6\right\rangle =\frac{65332016461837125}{1099511627776} && \left\langle \Theta ,\tau _5 \tau _6\right\rangle =\frac{24083995573458045225}{35184372088832} \\ & \left\langle \Theta ,\tau _6^2\right\rangle =\frac{1113215803724028329325}{140737488355328} && \left\langle \Theta ,\tau _1 \tau _7\right\rangle =\frac{297111189375}{2147483648} \\ & \left\langle \Theta ,\tau _2 \tau _7\right\rangle =\frac{162992299845375}{68719476736} && \left\langle \Theta ,\tau _3 \tau _7\right\rangle =\frac{9799801500864375}{274877906944} \\ & \left\langle \Theta ,\tau _4 \tau _7\right\rangle =\frac{17661596600472900075}{35184372088832} && \left\langle \Theta ,\tau _5 \tau _7\right\rangle =\frac{482393514590137475325}{70368744177664} \\ & \left\langle \Theta ,\tau _6 \tau _7\right\rangle =\frac{208660146935538633159825}{2251799813685248} && \left\langle \Theta ,\tau _7^2\right\rangle =\frac{11308033774288501710334875}{9007199254740992} \\ & \left\langle \Theta ,\tau _1 \tau _8\right\rangle =\frac{191751503518575}{274877906944} && \left\langle \Theta ,\tau _2 \tau _8\right\rangle =\frac{32281672904105625}{2199023255552} \\ & \left\langle \Theta ,\tau _3 \tau _8\right\rangle =\frac{18700513107631029675}{70368744177664} && \left\langle \Theta ,\tau _4 \tau _8\right\rangle =\frac{2497095829689640103925}{562949953421312} \\ & \left\langle \Theta ,\tau _5 \tau _8\right\rangle =\frac{638254566833734863087075}{9007199254740992} && \left\langle \Theta ,\tau _6 \tau _8\right\rangle =\frac{79821414874365136596248625}{72057594037927936} \\ & \left\langle \Theta ,\tau _7 \tau _8\right\rangle =\frac{158520492299731872217358075625}{9223372036854775808} && \left\langle \Theta ,\tau _8^2\right\rangle =\frac{9855445464368396327121143081625}{36893488147419103232} \\ & \left\langle \Theta ,\tau _1 \tau _9\right\rangle =\frac{4247525411254125}{1099511627776} && \left\langle \Theta ,\tau _2 \tau _9\right\rangle =\frac{6889659417119504025}{70368744177664} \\ & \left\langle \Theta ,\tau _3 \tau _9\right\rangle =\frac{295708708883846082825}{140737488355328} && \left\langle \Theta ,\tau _4 \tau _9\right\rangle =\frac{369515801101139991473175}{9007199254740992} \\ & \left\langle \Theta ,\tau _5 \tau _9\right\rangle =\frac{27307326135936642415995375}{36028797018963968} && \left\langle \Theta ,\tau _6 \tau _9\right\rangle =\frac{125147757076179666975625854375}{9223372036854775808} \\ & \left\langle \Theta ,\tau _7 \tau _9\right\rangle =\frac{1102253769039087864679419064125}{4611686018427387904} && \left\langle \Theta ,\tau _8 \tau _9\right\rangle =\frac{2470955043780035852615484506222625}{590295810358705651712} \\ & \left\langle \Theta ,\tau _9^2\right\rangle =\frac{173346999994671233640488824722852375}{2361183241434822606848} && \left\langle \Theta ,\tau _1 \tau _{10}\right\rangle =\frac{1640377818582027525}{70368744177664} \\ & \left\langle \Theta ,\tau _2 \tau _{10}\right\rangle =\frac{197139081099587301675}{281474976710656} && \left\langle \Theta ,\tau _3 \tau _{10}\right\rangle =\frac{79181956244767665764475}{4503599627370496} \\ & \left\langle \Theta ,\tau _4 \tau _{10}\right\rangle =\frac{28607674970974662230542875}{72057594037927936} && \left\langle \Theta ,\tau _5 \tau _{10}\right\rangle =\frac{77472421040242962853142648625}{9223372036854775808} \\ & \left\langle \Theta ,\tau _6 \tau _{10}\right\rangle =\frac{1574648241469678670322551398875}{9223372036854775808} && \left\langle \Theta ,\tau _7 \tau _{10}\right\rangle =\frac{500074235050199763259348761844125}{147573952589676412928} \\ & \left\langle \Theta ,\tau _8 \tau _{10}\right\rangle =\frac{313675523799849533828112798529713375}{4722366482869645213696} && \left\langle \Theta ,\tau _9 \tau _{10}\right\rangle =\frac{195855936811697982260208902271192680625}{151115727451828646838272} \\ & \left\langle \Theta ,\tau _{10}^2\right\rangle =\frac{15304091806682856653605975519597917118125}{604462909807314587353088}
\end{align*}
\end{tiny}

\paragraph*{Three-point intersection numbers, $1\leq \ell_1\leq \ell_2\leq\ell_3\leq 7$}

\begin{tiny}
\begin{align*}
&\left\langle \Theta ,\tau _1^3\right\rangle =\frac{7221}{2048}&& \left\langle \Theta ,\tau _1^2 \tau _2\right\rangle =\frac{524925}{32768} &&&\left\langle \Theta ,\tau _1 \tau _2^2\right\rangle =\frac{55787625}{524288} \\ &\left\langle \Theta ,\tau _2^3\right\rangle =\frac{8160299505}{8388608} && \left\langle \Theta ,\tau _1^2 \tau _3\right\rangle =\frac{19922175}{262144} &&& \left\langle \Theta ,\tau _1 \tau _2 \tau _3\right\rangle =\frac{2914222815}{4194304} \\ & \left\langle \Theta ,\tau _2^2 \tau _3\right\rangle =\frac{561519776475}{67108864}&& \left\langle \Theta ,\tau _1 \tau _3^2\right\rangle =\frac{200535367725}{33554432} &&& \left\langle \Theta ,\tau _2 \tau _3^2\right\rangle =\frac{49229655148485}{536870912} \\ & \left\langle \Theta ,\tau _3^3\right\rangle =\frac{5357097499513095}{4294967296} && \left\langle \Theta ,\tau _1^2 \tau _4\right\rangle =\frac{3237810975}{8388608}&&& \left\langle \Theta ,\tau _1 \tau _2 \tau _4\right\rangle =\frac{623885820075}{134217728} \\
 & \left\langle \Theta ,\tau _2^2 \tau _4\right\rangle =\frac{153158674747995}{2147483648}&& \left\langle \Theta ,\tau _1 \tau _3 \tau _4\right\rangle =\frac{54698188012965}{1073741824} &&& \left\langle \Theta ,\tau _2 \tau _3 \tau _4\right\rangle =\frac{16666510065902865}{17179869184} \\ & \left\langle \Theta ,\tau _3^2 \tau _4\right\rangle =\frac{2204149022466054615}{137438953472} && \left\langle \Theta ,\tau _1 \tau _4^2\right\rangle =\frac{18518016575263905}{34359738368} &&& \left\langle \Theta ,\tau _2 \tau _4^2\right\rangle =\frac{6857348740424943705}{549755813888} \\  
 & \left\langle \Theta ,\tau _3 \tau _4^2\right\rangle =\frac{1083235806125607211875}{4398046511104}&& \left\langle \Theta ,\tau _4^3\right\rangle =\frac{626729323148283152077875}{140737488355328} &&& \left\langle \Theta ,\tau _1^2 \tau _5\right\rangle =\frac{141786313515}{67108864} \\ & \left\langle \Theta ,\tau _1 \tau _2 \tau _5\right\rangle =\frac{34807868819955}{1073741824} && \left\langle \Theta ,\tau _2^2 \tau _5\right\rangle =\frac{10605949781451255}{17179869184} &&& \left\langle \Theta ,\tau _1 \tau _3 \tau _5\right\rangle =\frac{3787775648592705}{8589934592} \\
 & \left\langle \Theta ,\tau _2 \tau _3 \tau _5\right\rangle =\frac{1402639433346887505}{137438953472}&& \left\langle \Theta ,\tau _3^2 \tau _5\right\rangle =\frac{221570953666202985075}{1099511627776} &&& \left\langle \Theta ,\tau _1 \tau _4 \tau _5\right\rangle =\frac{1558468935931532625}{274877906944} \\ & \left\langle \Theta ,\tau _2 \tau _4 \tau _5\right\rangle =\frac{689331622581763917525}{4398046511104} && \left\langle \Theta ,\tau _3 \tau _4 \tau _5\right\rangle =\frac{128194632176429424912075}{35184372088832} &&& \left\langle \Theta ,\tau _4^2 \tau _5\right\rangle =\frac{86249350732236769967464575}{1125899906842624} \\ 
 \end{align*}
\begin{align*}
& \left\langle \Theta ,\tau _1 \tau _5^2\right\rangle =\frac{156664875383937753525}{2199023255552}&& \left\langle \Theta ,\tau _2 \tau _5^2\right\rangle =\frac{81578383422586747544925}{35184372088832} \\ & \left\langle \Theta ,\tau _3 \tau _5^2\right\rangle =\frac{17641912485909060186227775}{281474976710656}&& \left\langle \Theta ,\tau _4 \tau _5^2\right\rangle =\frac{13657290700342362804270453375}{9007199254740992} \\ & \left\langle \Theta ,\tau _5^3\right\rangle =\frac{2465542659153253894620947800875}{72057594037927936}&& \left\langle \Theta ,\tau _1^2 \tau _6\right\rangle =\frac{13387279450545}{1073741824} \\ & \left\langle \Theta ,\tau _1 \tau _2 \tau _6\right\rangle =\frac{4079138420722365}{17179869184}&& \left\langle \Theta ,\tau _2^2 \tau _6\right\rangle =\frac{1510533831861819765}{274877906944} \\ & \left\langle \Theta ,\tau _1 \tau _3 \tau _6\right\rangle =\frac{539469976573402875}{137438953472}&& \left\langle \Theta ,\tau _2 \tau _3 \tau _6\right\rangle =\frac{238614783964018637175}{2199023255552} \\ & \left\langle \Theta ,\tau _3^2 \tau _6\right\rangle =\frac{44375064151791685937625}{17592186044416}&& \left\langle \Theta ,\tau _1 \tau _4 \tau _6\right\rangle =\frac{265125172308079458375}{4398046511104} \\ & \left\langle \Theta ,\tau _2 \tau _4 \tau _6\right\rangle =\frac{138055725468255073854375}{70368744177664}&& \left\langle \Theta ,\tau _3 \tau _4 \tau _6\right\rangle =\frac{29855544173075376776945925}{562949953421312} \\ & \left\langle \Theta ,\tau _4^2 \tau _6\right\rangle =\frac{23112338094913132221232801725}{18014398509481984}&& \left\langle \Theta ,\tau _1 \tau _5 \tau _6\right\rangle =\frac{31376094449743999130175}{35184372088832} \\ & \left\langle \Theta ,\tau _2 \tau _5 \tau _6\right\rangle =\frac{18998979810187047955359075}{562949953421312}&& \left\langle \Theta ,\tau _3 \tau _5 \tau _6\right\rangle =\frac{4727523675542296407386332725}{4503599627370496} \\ & \left\langle \Theta ,\tau _4 \tau _5 \tau _6\right\rangle =\frac{4172456806519716753635289317625}{144115188075855872}&& \left\langle \Theta ,\tau _5^2 \tau _6\right\rangle =\frac{851876598415598596423734875348625}{1152921504606846976} \\ & \left\langle \Theta ,\tau _1 \tau _6^2\right\rangle =\frac{7307263707257464770845025}{562949953421312}&& \left\langle \Theta ,\tau _2 \tau _6^2\right\rangle =\frac{5091178832597722958533665225}{9007199254740992} \\ & \left\langle \Theta ,\tau _3 \tau _6^2\right\rangle =\frac{1444311966314562236238071599875}{72057594037927936}&& \left\langle \Theta ,\tau _4 \tau _6^2\right\rangle =\frac{1441637320153730808541734117691875}{2305843009213693952} \\ & \left\langle \Theta ,\tau _5 \tau _6^2\right\rangle =\frac{330507426847927743563704256091765375}{18446744073709551616}&& \left\langle \Theta ,\tau _6^3\right\rangle =\frac{143076665085625524144439793856692206125}{295147905179352825856} \\
&\left\langle \Theta ,\tau _1^2 \tau _7\right\rangle =\frac{679844026236375}{8589934592}  && \left\langle \Theta ,\tau _1 \tau _2 \tau _7\right\rangle =\frac{251752492250634375}{137438953472} \\ &  \left\langle \Theta ,\tau _2^2 \tau _7\right\rangle =\frac{111353523842933046675}{2199023255552}  && \left\langle \Theta ,\tau _1 \tau _3 \tau _7\right\rangle =\frac{39768774301596064725}{1099511627776} \\ &  \left\langle \Theta ,\tau _2 \tau _3 \tau _7\right\rangle =\frac{20708358416454371788125}{17592186044416}  && \left\langle \Theta ,\tau _3^2 \tau _7\right\rangle =\frac{4478331578178112272993375}{140737488355328} \\ &  \left\langle \Theta ,\tau _1 \tau _4 \tau _7\right\rangle =\frac{23009135858789763808125}{35184372088832}  && \left\langle \Theta ,\tau _2 \tau _4 \tau _7\right\rangle =\frac{13932585174890990776793625}{562949953421312} \\ &  \left\langle \Theta ,\tau _3 \tau _4 \tau _7\right\rangle =\frac{3466850693038052513223405375}{4503599627370496}  && \left\langle \Theta ,\tau _4^2 \tau _7\right\rangle =\frac{3059801657033466012792328078875}{144115188075855872} \\ &  \left\langle \Theta ,\tau _1 \tau _5 \tau _7\right\rangle =\frac{3166480939027367663526825}{281474976710656}  && \left\langle \Theta ,\tau _2 \tau _5 \tau _7\right\rangle =\frac{2206177493899231641551387625}{4503599627370496} \\ &  \left\langle \Theta ,\tau _3 \tau _5 \tau _7\right\rangle =\frac{625868518706912740538673268875}{36028797018963968}  && \left\langle \Theta ,\tau _4 \tau _5 \tau _7\right\rangle =\frac{624709505385771325499965588477875}{1152921504606846976} \\ &  \left\langle \Theta ,\tau _5^2 \tau _7\right\rangle =\frac{143219884965802657033319792435316375}{9223372036854775808}  && \left\langle \Theta ,\tau _1 \tau _6 \tau _7\right\rangle =\frac{848526584313597166576073475}{4503599627370496} \\ &  \left\langle \Theta ,\tau _2 \tau _6 \tau _7\right\rangle =\frac{674012204909537444497987815375}{72057594037927936}  && \left\langle \Theta ,\tau _3 \tau _6 \tau _7\right\rangle =\frac{216245597543221969005552576932625}{576460752303423488} \\ &  \left\langle \Theta ,\tau _4 \tau _6 \tau _7\right\rangle =\frac{242372112996098578290795791186518125}{18446744073709551616}  && \left\langle \Theta ,\tau _5 \tau _6 \tau _7\right\rangle =\frac{61999888203404304244773719191475446125}{147573952589676412928} \\ &  \left\langle \Theta ,\tau _6^2 \tau _7\right\rangle =\frac{29777660591694478072272815949606865930875}{2361183241434822606848}  && \left\langle \Theta ,\tau _1 \tau _7^2\right\rangle =\frac{112335035527164091943704528125}{36028797018963968} \\ &  \left\langle \Theta ,\tau _2 \tau _7^2\right\rangle =\frac{100914607388283481934842002427125}{576460752303423488}  && \left\langle \Theta ,\tau _3 \tau _7^2\right\rangle =\frac{36355816898739808749602039446104375}{4611686018427387904} \\ &  \left\langle \Theta ,\tau _4 \tau _7^2\right\rangle =\frac{45466584679729968624135916704886206375}{147573952589676412928}  && \left\langle \Theta ,\tau _5 \tau _7^2\right\rangle =\frac{12903652923026817062431054523642958979875}{1180591620717411303424} \\ &  \left\langle \Theta ,\tau _6 \tau _7^2\right\rangle =\frac{6840650520602244732461304449830082167319625}{18889465931478580854784}  && \left\langle \Theta ,\tau _7^3\right\rangle =\frac{1726520707483209249055570621199004902786559375}{151115727451828646838272}
\end{align*}
\end{tiny}

\paragraph*{Four-point intersection numbers, $1\leq\ell_1\leq\ell_2\leq\ell_3\leq\ell_4\leq 4$}

\begin{tiny}
\begin{align*}
&\left\langle \Theta ,\tau _1^4\right\rangle =\frac{4825971}{16384}&& \left\langle \Theta ,\tau _1^3 \tau _2\right\rangle =\frac{605705625}{262144} \\ & \left\langle \Theta ,\tau _1^2 \tau _2^2\right\rangle =\frac{102180197475}{4194304}&& \left\langle \Theta ,\tau _1 \tau _2^3\right\rangle =\frac{22305336602625}{67108864} \\ & \left\langle \Theta ,\tau _2^4\right\rangle =\frac{6118287865593075}{1073741824}&& \left\langle \Theta ,\tau _1^3 \tau _3\right\rangle =\frac{36491129325}{2097152} \\ & \left\langle \Theta ,\tau _1^2 \tau _2 \tau _3\right\rangle =\frac{7965945717975}{33554432}&& \left\langle \Theta ,\tau _1 \tau _2^2 \tau _3\right\rangle =\frac{2185058394718605}{536870912} \\ & \left\langle \Theta ,\tau _2^3 \tau _3\right\rangle =\frac{735717887208021375}{8589934592}&& \left\langle \Theta ,\tau _1^2 \tau _3^2\right\rangle =\frac{780361916123475}{268435456} \\ & \left\langle \Theta ,\tau _1 \tau _2 \tau _3^2\right\rangle =\frac{262752685378695225}{4294967296}&& \left\langle \Theta ,\tau _2^2 \tau _3^2\right\rangle =\frac{106548967987835464035}{68719476736} \\ & \left\langle \Theta ,\tau _1 \tau _3^3\right\rangle =\frac{38052819991224205965}{34359738368}&& \left\langle \Theta ,\tau _2 \tau _3^3\right\rangle =\frac{18292612579971274053495}{549755813888} \\ & \left\langle \Theta ,\tau _3^4\right\rangle =\frac{3673662570422147820860595}{4398046511104}&& \left\langle \Theta ,\tau _1^3 \tau _4\right\rangle =\frac{8850749243175}{67108864} \\ & \left\langle \Theta ,\tau _1^2 \tau _2 \tau _4\right\rangle =\frac{2427789302585325}{1073741824}&& \left\langle \Theta ,\tau _1 \tau _2^2 \tau _4\right\rangle =\frac{817452184156462575}{17179869184} \\ & \left\langle \Theta ,\tau _2^3 \tau _4\right\rangle =\frac{331485533529675544845}{274877906944}&& \left\langle \Theta ,\tau _1^2 \tau _3 \tau _4\right\rangle =\frac{291943042036776585}{8589934592} \\ & \left\langle \Theta ,\tau _1 \tau _2 \tau _3 \tau _4\right\rangle =\frac{118386498222267325155}{137438953472}&& \left\langle \Theta ,\tau _2^2 \tau _3 \tau _4\right\rangle =\frac{56910334852393854391665}{2199023255552} \\ & \left\langle \Theta ,\tau _1 \tau _3^2 \tau _4\right\rangle =\frac{20324969779924668510855}{1099511627776}&& \left\langle \Theta ,\tau _2 \tau _3^2 \tau _4\right\rangle =\frac{11429170458388415369176365}{17592186044416} \\ & \left\langle \Theta ,\tau _3^3 \tau _4\right\rangle =\frac{2654517578525246500814334825}{140737488355328}&& \left\langle \Theta ,\tau _1^2 \tau _4^2\right\rangle =\frac{131539156256344231395}{274877906944} \\ & \left\langle \Theta ,\tau _1 \tau _2 \tau _4^2\right\rangle =\frac{63233221816668147658785}{4398046511104}&& \left\langle \Theta ,\tau _2^2 \tau _4^2\right\rangle =\frac{35557413020470855924847955}{70368744177664} \\ & \left\langle \Theta ,\tau _1 \tau _3 \tau _4^2\right\rangle =\frac{12699006033434669177410125}{35184372088832}&& \left\langle \Theta ,\tau _2 \tau _3 \tau _4^2\right\rangle =\frac{8258498185220417475372945375}{562949953421312} \\ & \left\langle \Theta ,\tau _3^2 \tau _4^2\right\rangle =\frac{1125163811582917554083844364188525}{4611686018427387904}&& \left\langle \Theta ,\tau _1 \tau _4^3\right\rangle =\frac{9176069448469610909455503375}{1125899906842624} \\ & \left\langle \Theta ,\tau _2 \tau _4^3\right\rangle =\frac{875127272791496314312666747195875}{4611686018427387904}&& \left\langle \Theta ,\tau _3 \tau _4^3\right\rangle =\frac{332032047575230771772838453629775}{36893488147419103232} \\ & \left\langle \Theta ,\tau _4^4\right\rangle =\frac{1468690879523188482162010010875275}{590295810358705651712}
\end{align*}
\end{tiny}

\section{Tables of $n$-point correlators for $\nu\not=0$ (n=2,3,4)}\label{tablenu}

For an increasing sequence of indexes $0\leq \ell_1\leq \cdots\leq \ell_n$, introduce the notation
\be
\langle\Theta,\tau_{\ell_1}\cdots\tau_{\ell_n}\rangle_\nu:=\frac{2^{2\ell_1+1}\cdots 2^{2\ell_{n}+1}}{(2\ell_1+1)!!\cdots(2\ell_{n}+1)!!}\frac{1}{\left(\frac 1 2 -\nu\right)_{\ell_n+1}\left(\frac 1 2 +\nu\right)_{\ell_n+1}}\left.\frac{\pa \log\tau(\t;\nu)}{\pa t_{\ell_1}\cdots \pa t_{\ell_n}}\right|_{\t=0}.
\ee 
Note that
\be
\left.\langle\Theta,\tau_{\ell_1}\cdots\tau_{\ell_n}\rangle_\nu\right|_{\nu=0}=\frac{2^{2\ell_n+2}}{(2\ell_n+1)!!^2}\langle\Theta,\tau_{\ell_1}\cdots\tau_{\ell_n}\rangle.
\ee
Below we list some correlators $\langle\Theta,\tau_{\ell_1}\cdots\tau_{\ell_n}\rangle_\nu$ for $n=2,3,4$ and $1\leq \ell_1\leq\cdots \leq \ell_n$; insertions of arbitrary positive powers $\tau_0$ are not considered, as the corresponding correlators can be computed from the relations
\be 
\langle\Theta,\tau_0^k\tau_{\ell_1}\cdots\tau_{\ell_n}\rangle_\nu=\left(n+2\sum_{i=1}^n\ell_i\right)_k\langle\Theta,\tau_{\ell_1}\cdots\tau_{\ell_n}\rangle_\nu, \qquad
\langle\Theta,\tau_0\rangle_\nu=\frac 12
\ee
which follow from the Virasoro constraint $L_0\tau=0$.

\paragraph*{Two-point correlators, $1\leq \ell_1\leq \ell_2\leq 7$}

\begin{tiny}
\begin{align*}
&\left\langle \Theta ,\tau _1^2\right\rangle_\nu =\frac{21-4 \nu ^2}{96}\qquad
\left\langle \Theta ,\tau _1 \tau _2\right\rangle_\nu =\frac{115-12 \nu ^2}{1536}\qquad
\left\langle  \Theta ,\tau _2^2\right\rangle_\nu =\frac{48 \nu ^4-1240 \nu ^2+8371}{30720}\qquad
\left\langle \Theta , \tau _1 \tau _3\right\rangle_\nu =\frac{61-4 \nu ^2}{3840}\\ 
&\left\langle \Theta ,\tau _2 \tau _3\right\rangle_\nu =\frac{16 \nu ^4-616 \nu ^2+6489}{73728}\qquad
\left\langle \Theta ,\tau _3^2\right\rangle_\nu =\frac{-320 \nu ^6+22960 \nu ^4-587804 \nu ^2+5087601}{10321920}\qquad
\left\langle \Theta ,\tau _1 \tau _4\right\rangle_\nu =\frac{89-4 \nu ^2}{36864}\\ 
&\left\langle \Theta ,\tau _2 \tau _4\right\rangle_\nu =\frac{240 \nu ^4-12920 \nu ^2+195407}{10321920}\qquad
\left\langle \Theta ,\tau _3 \tau _4\right\rangle_\nu =\frac{-320 \nu ^6+30960 \nu ^4-1100604 \nu ^2+13452101}{94371840}\\ 
&\left\langle \Theta ,\tau _4^2\right\rangle_\nu =\frac{1280 \nu ^8-195840 \nu ^6+12179424 \nu ^4-345644240 \nu ^2+3670308261}{3397386240}\qquad
\left\langle \Theta ,\tau _1 \tau _5\right\rangle_\nu =\frac{367-12 \nu ^2}{1290240}\\ 
&\left\langle \Theta ,\tau _2 \tau _5\right\rangle_\nu =\frac{48 \nu ^4-3448 \nu ^2+70747}{23592960}\qquad
\left\langle \Theta ,\tau _3 \tau _5\right\rangle_\nu =\frac{-64 \nu ^6+8048 \nu ^4-379180 \nu ^2+6204501}{212336640}\\ 
&\left\langle \Theta ,\tau _4 \tau _5\right\rangle_\nu =\frac{5376 \nu ^8-1044736 \nu ^6+84295904 \nu ^4-3137766544 \nu ^2+44120931525}{158544691200}\\ &\left\langle \Theta ,\tau _5^2\right\rangle_\nu =\frac{-21504 \nu ^{10}+6012160 \nu ^8-734439552 \nu ^6+46399124640 \nu ^4-1474066134244 \nu ^2+18569159714025}{6975966412800}\\ 
&\left\langle \Theta ,\tau _1 \tau _6\right\rangle_\nu =\frac{161-4 \nu ^2}{5898240}\qquad\left\langle \Theta ,\tau _2 \tau _6\right\rangle_\nu =\frac{16 \nu ^4-1480 \nu ^2+39537}{106168320}\qquad 
\left\langle \Theta ,\tau _3 \tau _6\right\rangle_\nu =\frac{-448 \nu ^6+71120 \nu ^4-4287892 \nu ^2+90370575}{19818086400}\\
&\left\langle \Theta ,\tau _4 \tau _6\right\rangle_\nu =\frac{1792 \nu ^8-431872 \nu ^6+43883168 \nu ^4-2071941488 \nu ^2+37189031175}{697596641280}\\ 
&\left\langle \Theta ,\tau _5 \tau _6\right\rangle_\nu =\frac{-7168 \nu ^{10}+2446080 \nu ^8-370790784 \nu ^6+29295092320 \nu ^4-1171373444748 \nu ^2+18694588685175}{30440580710400}\\ &\left\langle \Theta ,\tau _6^2\right\rangle_\nu =\frac{28672 \nu ^{12}-13232128 \nu ^{10}+2793912576 \nu ^8-327025863424 \nu ^6+21768252203152 \nu ^4-770335337110248 \nu ^2+11233370707313175}{1582910196940800}\\ 
&\left\langle \Theta ,\tau _1 \tau _7\right\rangle_\nu =\frac{205-4 \nu ^2}{92897280}\qquad
\left\langle \Theta ,\tau _2 \tau _7\right\rangle_\nu =\frac{48 \nu ^4-5560 \nu ^2+187435}{4954521600}\qquad
\left\langle \Theta ,\tau _3 \tau _7\right\rangle_\nu =-\frac{\left(4 \nu ^2-229\right) \left(16 \nu ^4-2216 \nu ^2+108265\right)}{43599790080}\\ 
&\left\langle \Theta ,\tau _4 \tau _7\right\rangle_\nu =\frac{1280 \nu ^8-375040 \nu ^6+46863584 \nu ^4-2734598160 \nu ^2+60930510741}{7610145177600}\\ 
&\left\langle \Theta ,\tau _5 \tau _7\right\rangle_\nu =\frac{-3072 \nu ^{10}+1258240 \nu ^8-231836544 \nu ^6+22386337632 \nu ^4-1098992901244 \nu ^2+21634639864743}{197863774617600}\\ 
&\left\langle \Theta ,\tau _6 \tau _7\right\rangle_\nu =\frac{4096 \nu ^{12}-2242560 \nu ^{10}+569358080 \nu ^8-80608549120 \nu ^6+6520060384752 \nu ^4-281678271771320 \nu ^2+5038977351919497}{3409345039564800}\\ 
&\left\langle \Theta ,\tau _7^2\right\rangle_\nu =(-16384 \nu ^{14}+11612160 \nu ^{12}-3887657984 \nu ^{10}+754214844160 \nu ^8-89084725490880 \nu ^6\\&\phantom{{}\left\langle \Theta ,\tau _7^2\right\rangle_\nu =}+6317860403726480 \nu ^4-247182521760945852 \nu ^2+4096200945908249325)/{204560702373888000}
\end{align*}
\end{tiny}

\paragraph*{Three-point correlators, $1\leq \ell_1\leq \ell_2\leq\ell_3\leq 4$}

\begin{tiny}
\begin{align*}
&\left\langle \Theta ,\tau _1^3\right\rangle_\nu=\frac{1}{384} \left(4 \nu ^2-29\right) \left(12 \nu ^2-83\right) 
\qquad\left\langle \Theta ,\tau _1^2 \tau _2\right\rangle_\nu=\frac{1}{512} \left(16 \nu ^4-376 \nu ^2+2333\right)
\qquad\left\langle \Theta ,\tau _1 \tau _2^2\right\rangle_\nu=\frac{-192 \nu ^6+8720 \nu ^4-138980 \nu ^2+743835}{24576}
\\&\left\langle \Theta ,\tau _2^3\right\rangle_\nu=\frac{3840 \nu ^8-285440 \nu ^6+8415904 \nu ^4-111717680 \nu ^2+544019967}{1966080}
\qquad\left\langle \Theta ,\tau _1^2 \tau _3\right\rangle_\nu=\frac{16 \nu ^4-568 \nu ^2+5421}{3072}
\\&\left\langle \Theta ,\tau _1 \tau _2 \tau _3\right\rangle_\nu=\frac{-960 \nu ^6+62480 \nu ^4-1468628 \nu ^2+11894787}{737280}
\qquad\left\langle \Theta ,\tau _2^2 \tau _3\right\rangle_\nu=\frac{\left(4 \nu ^2-97\right) \left(192 \nu ^6-14992 \nu ^4+455716 \nu ^2-4725603\right)}{2359296}
\\&\left\langle \Theta ,\tau _1 \tau _3^2\right\rangle_\nu=\frac{\left(4 \nu ^2-97\right) \left(64 \nu ^6-5232 \nu ^4+162476 \nu^2-1687653\right)}{1179648}
\\&\left\langle \Theta ,\tau _2 \tau _3^2\right\rangle_\nu=\frac{-5120 \nu ^{10}+779520 \nu ^8-51300480 \nu ^6+1748059040 \nu ^4-29897734692 \nu ^2+200937367953}{94371840}
\\&\left\langle \Theta ,\tau _3^3\right\rangle_\nu=\frac{143360 \nu ^{12}-30464000 \nu ^{10}+2915754240 \nu ^8-154331121920 \nu ^6+4618556633936 \nu ^4-72493109900568 \nu ^2+459179785672551}{15854469120}
\\& \left\langle \Theta ,\tau _1^2 \tau _4\right\rangle_\nu=\frac{48 \nu ^4-2408 \nu ^2+32631}{73728}
\qquad \left\langle \Theta ,\tau _1 \tau _2 \tau _4\right\rangle_\nu=\frac{-192 \nu ^6+17040 \nu ^4-554020 \nu ^2+6287587}{1179648}
\\&\left\langle \Theta ,\tau _2^2 \tau _4\right\rangle_\nu=\frac{3840 \nu ^8-520960 \nu ^6+29220000 \nu ^4-767152560 \nu ^2+7717746271}{94371840}
\\&\left\langle \Theta ,\tau _1 \tau _3 \tau _4\right\rangle_\nu=\frac{1280 \nu ^8-179200 \nu ^6+10299168 \nu ^4-273679520 \nu ^2+2756270497}{47185920}
\\&\left\langle \Theta ,\tau _2 \tau _3 \tau _4\right\rangle_\nu=\frac{-107520 \nu ^{10}+21136640 \nu ^8-1826659968 \nu ^6+82857551520 \nu ^4-1913355449780 \nu ^2+17636518588257}{15854469120} 
\\&\left\langle \Theta ,\tau _3^2 \tau _4\right\rangle_\nu=\frac{20480 \nu ^{12}-5498880 \nu ^{10}+676050688 \nu ^8-46592384768 \nu ^6+1840104258096 \nu ^4-38670430868392 \nu ^2+333204689715201}{18119393280}
\\&\left\langle \Theta ,\tau _1 \tau _4^2\right\rangle_\nu=\frac{-107520 \nu ^{10}+22140160 \nu ^8-1976843904 \nu ^6+91322556576 \nu ^4-2124380314036 \nu ^2+19595784735729}{31708938240} 
\\&\left\langle \Theta ,\tau _2 \tau _4^2\right\rangle_\nu=\frac{61440 \nu ^{12}-16803840 \nu ^{10}+2088837376 \nu ^8-144682986240 \nu ^6+5723155068432 \nu ^4-120305338397800 \nu ^2+1036636241938767}{72477573120}
\\&\left\langle \Theta ,\tau _3 \tau _4^2\right\rangle_\nu=(-81920 \nu ^{14}+29306880 \nu ^{12}-4916978688 \nu ^{10}+480717922048 \nu ^8-28705415560128 \nu ^6
\\ &\phantom{{}\left\langle \Theta ,\tau _3 \tau _4^2\right\rangle_\nu=}+1026041519901072 \nu ^4-20052853905009164 \nu ^2+163754468046199125)/579820584960
\\&\left\langle \Theta ,\tau _4^3\right\rangle_\nu=(6881280 \nu ^{16}-3169976320 \nu ^{14}+698577567744 \nu ^{12}-92301918593024 \nu ^{10}+7763986997949952 \nu ^8-417263450232233472 \nu ^6\\
&\phantom{{}\left\langle \Theta ,\tau _4^3\right\rangle_\nu=}+13803637161401407424 \nu ^4-254559676442493789984 \nu ^2+1989616898883438578025)/389639433093120
\end{align*}
\end{tiny}

\paragraph*{Four-point intersection numbers, $1\leq \ell_1\leq \ell_2\leq\ell_3\leq \ell_4\leq 3$}

\begin{tiny}
\begin{align*}
&\left\langle \Theta ,\tau _1^4\right\rangle_\nu =\frac{-704 \nu ^6+19216 \nu ^4-178436 \nu ^2+536219}{1024} \\ 
&\left\langle \Theta ,\tau _1^3 \tau _2\right\rangle_\nu =\frac{-832 \nu ^6+35120 \nu ^4-526588 \nu ^2+2692025}{4096} \\ 
&\left\langle \Theta ,\tau _1^2 \tau _2^2\right\rangle_\nu =\frac{11520 \nu ^8-809984 \nu ^6+22719008 \nu ^4-289118880 \nu ^2+1362402633}{196608} \\ 
&\left\langle \Theta ,\tau _1 \tau _2^3\right\rangle_\nu =\frac{-52224 \nu ^{10}+5533440 \nu ^8-250503040 \nu ^6+5814287840 \nu ^4-66908033020 \nu ^2+297404488035}{3145728} \\ 
&\left\langle \Theta ,\tau _2^4\right\rangle_\nu =\frac{1167360 \nu ^{12}-174766080 \nu ^{10}+11710001920 \nu ^8-431798964480 \nu ^6+8939142476592 \nu ^4-95917055510200 \nu ^2+407885857706205}{251658240} \\ 
&\left\langle \Theta ,\tau _1^3 \tau _3\right\rangle_\nu =\frac{-320 \nu ^6+19568 \nu ^4-431484 \nu ^2+3309853}{8192} \\ 
&\left\langle \Theta ,\tau _1^2 \tau _2 \tau _3\right\rangle_\nu =\frac{13056 \nu ^8-1273344 \nu ^6+50595680 \nu ^4-931233344 \nu ^2+6502812831}{1179648} \\ 
&\left\langle \Theta ,\tau _1 \tau _2^2 \tau _3\right\rangle_\nu =\frac{-291840 \nu ^{10}+41463040 \nu ^8-2569249920 \nu ^6+83271019680 \nu ^4-1368420867076 \nu ^2+8918605692729}{94371840} \\ 
&\left\langle \Theta ,\tau _2^3 \tau _3\right\rangle_\nu =\frac{86016 \nu ^{12}-16799744 \nu ^{10}+1496199936 \nu ^8-74679477504 \nu ^6+2135232115376 \nu ^4-32393359813080 \nu ^2+200195343457965}{100663296} \\ 
&\left\langle \Theta ,\tau _1^2 \tau _3^2\right\rangle_\nu =\frac{-19456 \nu ^{10}+2852608 \nu ^8-180524928 \nu ^6+5917711072 \nu ^4-97684465660 \nu ^2+637030135611}{9437184} \\ 
&\left\langle \Theta ,\tau _1 \tau _2 \tau _3^2\right\rangle_\nu =\frac{86016 \nu ^{12}-17262592 \nu ^{10}+1567786752 \nu ^8-79231405824 \nu ^6+2280610471216 \nu ^4-34693343884584 \nu ^2+214491988064241}{150994944} \\ 
&\left\langle \Theta ,\tau _2^2 \tau _3^2\right\rangle_\nu =(-5652480 \nu ^{14}+1492234240 \nu ^{12}-184371696640 \nu ^{10}+13285907930880 \nu ^8-585013383321280 \nu ^6\\
& \phantom{{}\left\langle \Theta ,\tau _2^2 \tau _3^2\right\rangle_\nu =}+15389216258794000 \nu ^4-220180441522580316 \nu ^2+1304681240667373029)/36238786560 \\ 
&\left\langle \Theta ,\tau _1 \tau _3^3\right\rangle_\nu =(-1884160 \nu ^{14}+509071360 \nu ^{12}-64051706880 \nu ^{10}+4675919560960 \nu ^8-207588275983936 \nu ^6\\
&\phantom{{}\left\langle \Theta ,\tau _1 \tau _3^3\right\rangle_\nu =}+5484821959054704 \nu ^4-78614370887838804 \nu ^2+465952897851724971)/18119393280 \\ 
&\left\langle \Theta ,\tau _2 \tau _3^3\right\rangle_\nu =(8192000 \nu ^{16}-2819031040 \nu ^{14}+463817768960 \nu ^{12}-45840708300800 \nu ^{10}+2888541257222656 \nu ^8-116204548461042944 \nu ^6\\
&\phantom{{}\left\langle \Theta ,\tau _2 \tau _3^3\right\rangle_\nu =}+2866333931776933632 \nu ^4-39127411662526409040 \nu ^2+223991174448627845553)/289910292480 \\ 
&\left\langle \Theta ,\tau _3^4\right\rangle_\nu =(-9175040 \nu ^{18}+4002611200 \nu ^{16}-849849548800 \nu ^{14}+110927057633280 \nu ^{12}-9520136272668672 \nu ^{10}+544814126675069440 \nu ^8\\
&\phantom{{}\left\langle \Theta ,\tau _3^4\right\rangle_\nu =}-20445062672058146560 \nu ^6+478943053050627574976 \nu ^4-6290239045745431301868 \nu ^2+34987262575449026865339)/1803886264320
\end{align*}
\end{tiny}

\bibliographystyle{alpha}

\newcommand{\etalchar}[1]{$^{#1}$}

\end{document}